\documentclass{elsart}

\usepackage{natbib}

\usepackage{epsfig}

\usepackage{amssymb}

\newcommand{\luerg}{erg s$^{-1}$ cm$^{-2}$ sr$^{-1}$}
\newcommand{\flux}{erg s$^{-1}$ cm$^{-2}$}
\newcommand{\ncm}{cm$^{-3}$}
\newcommand{\ncmK}{cm$^{-3}$ K}
\newcommand{\Ncm}{cm$^{-2}$}
\newcommand{\kms}{km s$^{-1}$}
\newcommand{\um}{$\mu$m}

\newcommand{\Hi}{H {\small I}}

\newcommand{\hh}{H$_2$}
\newcommand{\HH}{H$_2$ $\upsilon=1\rightarrow0$ S(1)}

\newcommand{\Neii}{[Ne {\small II}]}

\newcommand{\Arii}{[Ar {\small II}]}
\newcommand{\Feii}{[Fe {\small II}]}
\newcommand{\CO}{$^{12}$CO $J=2\rightarrow1$}
\newcommand{\COv}{$^{12}$CO $v=1-0$}

\newcommand{\ionNeii}{Ne$^{+}$}

\newcommand{\ionArii}{Ar$^{+}$}
\newcommand{\ionFeii}{Fe$^{+}$}

\newcommand{\Xco}{$X_{\textrm{\tiny{CO}}}$}

\newcommand{\nhh}{$n(\textrm{H}_2)$}
\newcommand{\nh}{$n(\textrm{H})$}

\newcommand{\vs}{$\upsilon_s$}

\newcommand{\Nh}{$N(\textrm{H})$}
\newcommand{\Nhh}{$N(\textrm{H}_2)$}
\newcommand{\NhhIRC}{$N$(H$_{2}\,; T>100\,\textrm{K})$}

\newcommand{\akari}{\textit{AKARI}}
\newcommand{\spitzer}{\textit{Spitzer}}

\begin{document}

\begin{frontmatter}



\title{Infrared Studies of Molecular Shocks in the Supernova Remnant HB 21:\\ II. Thermal Admixture of Shocked \hh{} Gas in the South}


\author{Jong-Ho Shinn\corauthref{cor}}
\address{Dept. of Physics and Astronomy, FPRD, Seoul National University, 599 Gwanangno, Gwanak-gu, Seoul, 151-747, South Korea}
\corauth[cor]{Corresponding author}
\ead{jhshinn@snu.ac.kr}

\author{Bon-Chul Koo}
\address{Dept. of Physics and Astronomy, FPRD, Seoul National University, 599 Gwanangno, Gwanak-gu, Seoul, 151-747, South Korea}
\ead{koo@astrohi.snu.ac.kr}

\author{Michael Burton}
\address{School of Physics, University of New South Wales, Sydney, New South Wales 2052, Australia}
\ead{m.burton@unsw.edu.au}

\author{Ho-Gyu Lee}
\address{Dept. of Astronomy, Graduate School of Science, the University of Tokyo, 7-3-1 Hongo, Bunkyo-ku, Tokyo 113-0003, Japan}
\ead{hglee@astron.s.u-tokyo.ac.jp}

\author{Dae-Sik Moon}
\address{Dept. of Astronomy and Astrophysics, University of Toronto, Toronto, ON M5S 3H4, Canada}
\ead{moon@astro.utoronto.ca}

\begin{abstract}
We present near- and mid-infrared observations on the shock-cloud interaction region in the southern part of the supernova remnant HB 21, performed with the InfraRed Camera (IRC) aboard \akari{} satellite and the Wide InfraRed Camera (WIRC) at the Palomar 5 m telescope.
The IRC 4 \um{} (N4), 7 \um{} (S7), and 11 \um{} (S11) band images and the WIRC \HH{} 2.12 \um{} image show similar diffuse features, around a shocked CO cloud.
We analyzed the emission through comparison with the \hh{} line emission of several shock models.
The IRC colors are well explained by the thermal admixture model of \hh{} gas---whose infinitesimal \hh{} column density has a power-law relation with the temperature $T$, $dN\sim T^{-b}dT$---with \nhh{} $\sim3.9\times10^4$ \ncm, $b\sim4.2$, and \NhhIRC{} $\sim2.8\times10^{21}$ \Ncm.
We interpreted these parameters with several different pictures of the shock-cloud interactions---multiple planar C-shocks, bow shocks, and shocked clumps---and discuss their weaknesses and strengths.
The observed \HH{} intensity is four times greater than the prediction from the power-law admixture model, the same tendency as found in the northern part of HB 21 (Paper I).
We also explored the limitation of the thermal admixture model with respect to the derived model parameters.

\end{abstract}


\end{frontmatter}

\parindent=0.5 cm

\section{Introduction}
HB 21 (G89.0+4.7) is a large ($\sim120'\times90'$), middle-aged ($\sim5000-7000$ yr, \citealt{Lazendic(2006)ApJ_647_350,Byun(2006)ApJ_637_283}) supernova remnant (SNR) at a distance estimated to be from $\sim0.8$ kpc to $\sim1.7$ kpc \citep{Leahy(1987)MNRAS_228_907,Tatematsu(1990)A&A_237_189,Byun(2006)ApJ_637_283}.
Based on its indented, shell-like appearance in the radio and the existence of nearby giant molecular clouds, it is thought to be interacting with a molecular cloud \citep[cf.~Fig.~\ref{fig-obs};][]{Erkes(1969)AJ_74_840,Huang(1986)ApJ_309_804,Tatematsu(1990)A&A_237_189}.
The first direct evidence for this interaction was the detection of broad CO emission lines near the edge and the center of the remnant \citep{Koo(2001)ApJ_552_175,Byun(2006)ApJ_637_283}.
The existence of such an interaction was further supported by the suggestion that evaporation of the cloud might be responsible for the enhanced thermal X-rays seen in the central part of the remnant \citep{Leahy(1996)A&A_315_260}.

We performed infrared imaging observations toward two localized positions in HB 21, where the broad CO emission lines were observed (Fig.~\ref{fig-obs}), with two instruments: the InfraRed Camera \citep[IRC,][]{Onaka(2007)PASJ_59_S401s} aboard a Japanese satellite, \akari{} \citep{Murakami(2007)PASJ_59_S369s} and the Wide-field InfraRed Camera \citep[WIRC,][]{Wilson(2003)inproc} on the Palomar 5 m Hale telescope.
From the analysis of the northern part (``Cloud N'') data, we found that the mid-infrared diffuse features originated from shocked \hh{} gas, with their excitation conditions well described by a thermal admixture of \hh{} gas, whose infinitesimal \hh{} column density has a power-law relation with the temperature $T$, d$N\sim T^{-b}dT$ \citep[][hereafter Paper I]{Shinn(2009)ApJ_693_1883}.
Such \hh{} excitation conditions are consistent with the ``ankle-like'' energy level population diagram (i.e. a turn-up in population for higher energies, see Fig.~\ref{fig-pop}), hitherto observed at the shock-cloud interaction regions (cf. $\S$~1 of Paper I).

Here we present the analysis of the southern portion of HB 21 (``Cloud S''), following the method of Paper I.
The near- and mid-infrared images ($\sim2-13$ \um) we obtained show diffuse features around a shocked CO cloud.
We analyze them as emission lines of \hh{} gas in statistical equilibrium.
We find the emission, as with the Cloud N case, to be well described with a power-law admixture model of thermal \hh{} gas.
We then discuss these results with physical pictures of the shock-cloud interaction.

\section{Observations} \label{obs}
We observed two specific regions (Cloud N and Cloud S in Fig.~\ref{fig-obs}), where slow shocks ($\lesssim20$ \kms) propagate into clouds of \nhh{} $\sim10^3$ \ncm{} \citep{Koo(2001)ApJ_552_175}, using two different instruments: IRC \citep{Onaka(2007)PASJ_59_S401s} aboard the \akari{} satellite and WIRC \citep{Wilson(2003)inproc} on the Palomar 5 m telescope.
The Cloud N data were analyzed in Paper I, and the Cloud S data are analyzed here.
Details on the observations and reduction of the IRC and WIRC data are described separately, below.

\subsection{\akari{} IRC observations} \label{obs-irc}
\akari{} is a satellite designed for both imaging and spectroscopy in the infrared \citep{Murakami(2007)PASJ_59_S369s}.
The IRC is one of \akari{}'s scientific instruments, which covers the wavelength range 2--30 \um{} and has a $\sim10'\times10'$ field-of-view for imaging.
The IRC pointed-imaging observations for Cloud S were performed on 2007 Jun 3rd towards (RA, Dec) = ($20^h46^m07.80^s$, $+50^\circ02'02.00''$) in J2000.
IRC comprises three channels (NIR, MIR-S, and MIR-L), each of which has three band-pass filters for imaging.
Among these, we employed four filters from NIR and MIR-S channels for the observations; the MIR-L channel was not used for observing Cloud S, due to lack of observing time.
Table \ref{tbl-obs} lists the wavelength coverage and the imaging resolutions ($\Gamma$), together with pixel sizes in each channel.

Data reduction was the same as for the Cloud N data (cf. Paper I), except for flat-fielding; we used a different MIR-S flat, since the dark pattern seen in the channel changed around 2007 Jan 7th\footnote{This is described in a note at http://www.ir.isas.jaxa.jp/AKARI/Observation/DataReduction/IRC/}.
We obtained the refined coadded image through the IRC Imaging Pipeline \citep[v.~20070104][]{Lorente(2007)man}.
Astrometric information was added to the coadded images, employing the 2MASS catalog \citep{Skrutskie(2006)AJ_131_1163s}, with a matching tolerance of 1.5 pixels.
The systematic errors ($\sim2-5\%$) of the calibration were included in the error estimation, as done for the Cloud N data.
Then, for the comparison between images from different bands, the pixel size was interpolated to $1''$ and the spatial resolution was smoothed to $\simeq7.43''$.
Point sources were removed applying the DAOPHOT package \citep{Stetson(1987)PASP_99_191} of IRAF making use of the simple-masking method.
The final images for Cloud S are displayed in Figure~\ref{fig-result}.

\subsection{Palomar WIRC \hh{} observations} \label{obs-wirc}
The WIRC observations were taken together with those of Cloud N (cf. Paper I).
We carried out the \HH{} 2.12 \um{} narrow-band filter imaging observation of Cloud S, centered at (RA, Dec) = (20h:46m:23.28s, $+49^\circ:54':16.72''$) in J2000, on the Palomar 5 m Hale telescope on 2005 August 29.
The WIRC is equipped with a Rockwell Science Hawaii II HgCdTe 2K infrared focal plane array, covering a $8.7'\times8.7'$ field of view with a $\sim0.25''$ pixel scale.
The WIRC field partially covered the IRC field, due to the mislocation of the WIRC observation center (cf. Fig.~\ref{fig-result}).
The data reduction was also the same as that of Cloud N data.
50 dithered images of 30 sec exposure were obtained.
We subtracted dark and sky background from each individual dithered frame and then divided by a normalized flat frame.
Finally, the dithered frames were combined to produce the final image.

Astrometry was obtained by matching the positions of 13 field point-sources with those of 2MASS catalog sources, and the positions agreed within $\sim0.3''$.
Flux calibration was also done using the 2MASS catalog.
We matched the magnitudes of 13 field point-sources with the corresponding $K_s$ magnitudes from the 2MASS catalog.
Their correlation coefficient was 0.9956, and their ratio, $M_{WIRC}/M_{K_s}$, was $1.026\pm0.011$.
The systematic error ($\sim14\%$) of the calibration was included in the error estimation.
Point sources were removed using DAOPHOT.
The full-width-at-half-maximum (FWHM) of these sources was found to be $\sim1.1''$.

\section{Results} \label{res}
The final images of IRC and WIRC are displayed in Figure~\ref{fig-result}, together with a \CO{} 230.583 GHz \citep{Koo(2001)ApJ_552_175} image for reference.
The peak positions of the shocked S1 and S2 clouds, where broad \CO{} lines were observed \citep{Koo(2001)ApJ_552_175}, are also indicated on the images.

\subsection{Morphology} \label{res-mor}
The IRC images (Fig.~\ref{fig-result}) show different features from band to band, as in Cloud N (Paper I).
The N3 and N4 images are dominated by point sources, while the S7 and S11 images show similar diffuse features.
They do not, however, look like bow shocks or planar shocks, unlike in Cloud N.
Around the cloud S1, common diffuse features are seen in all the IRC images, although they are faint in the N3 image.
However, there are no such diffuse features around the cloud S2.
This is in contrast with Cloud N, where the shocked CO clouds have corresponding diffuse features in the IRC S7 and S11 bands; this is more interesting considering that the cloud properties observed from CO emission lines are similar for both clouds, S1 and S2 \citep{Koo(2001)ApJ_552_175}.

Higher extinction toward the cloud S2 than the cloud S1 does not seem to be the reason for the absence of diffuse IRC features around the cloud S2, since very high column density \Nh$\,\gtrsim10^{23}$ \Ncm{} is required for the extinction to be effective at $\sim$ 10 \um{} \citep{Draine(2003)ARA&A_41_241}.
Recalling that \hh{} emissions are the main source for the diffuse IRC features in the case of Cloud N (Paper I), this absence may be caused by the lack of \hh{} gas around the cloud S2. 
The dissociation of \hh{} by hot gas ($\gtrsim10^6$ K) can be the reason, however it is unlikely since the X-ray emission is not strong around the cloud S2 \citep{Byun(2006)ApJ_637_283}.
The X-ray flux ($0.1-2.4$ keV) of HB 21 is $31.8\times10^{-10}$ \flux{} \citep{Leahy(1996)A&A_315_260}, and the cloud S2 locates $\sim10'$ away from the central X-ray emissions \citep{Byun(2006)ApJ_637_283}, which corresponds to a projected distance of $2-5$ pc.
With a hydrogen nuclei density of $10^4$ \ncm{} and an attenuating column density of $10^{20}-10^{22}$ \Ncm, this distance corresponds to an effective ionization parameter, Log $\xi_{eff}$, ranging from $-4$ to $-7$, sufficiently small so that the effects of X-rays are negligible \citep{Maloney(1996)ApJ_466_561}.
At the moment, it remains uncertain why such an absence of diffuse IRC features happens \emph{only} to the cloud S2.

A diffuse, looplike feature with a diameter of $\sim4'$ is seen in the northern portion of the S7 and S11 images, however, it does not seem to be related with the shocked CO clouds S1 and S2.
The \CO{} map shows a similar looplike feature at 9.4 \kms{} \citep{Koo(2001)ApJ_552_175}.
Since the looplike feature looks similar in the IRC S7 and S11 bands, they may be generated by \Arii{} 6.99 \um{} and \Neii{} 12.8 \um{} emission lines, which are expected to show similar distributions in the shock-cloud interaction regions considering their ionization potentials \citep{Neufeld(2007)ApJ_664_890}; indeed, these lines have been frequently observed around SNRs \citep[e.g.~][]{Arendt(1999)ApJ_521_234,Oliva(1999)A&A_343_943,Reach(2002)ApJ_564_302,Neufeld(2007)ApJ_664_890}.
Thermal emission from warm dust ($\gtrsim100$ K) is another candidate.
However, it seems unlikely because hot gas ($\gtrsim10^6$ K)---the heat source for the warm dust---is not abundant around the cloud S1 \cite[Fig.~1 of][]{Byun(2006)ApJ_637_283}.

The WIRC \HH{} image only covered a small portion of the field of the IRC images because of the mislocation, but does include the cloud S1 (cf. Fig.~\ref{fig-result}).
The \hh{} image shows diffuse features around the cloud S1, similar to those seen in the IRC images.
The similarity is more easily recognizable in Figure \ref{fig-rgb}, which zooms into the area around the cloud S1.
The RGB color image is made with N4 (blue), S7 (green), and S11 (red).
Three elongated clumps, whose sizes are comparable with the FWHM of the IRC images ($\sim7.4''$), are apparent.
Overall, their colors are red-and-yellowish, although the southwestern part of the features shows a little bluish color.
The filamentary features seen in the WIRC image have a similar overall morphology to those seen in the RGB image.
They also surround the shocked CO cloud, S1 (cf.~Fig.~\ref{fig-result} and \ref{fig-rgb}).
This geometrical relationship thus suggests that the diffuse infrared features seen around the cloud S1 may also originate from shock excitation.

\subsection{Quantitative Infrared Characteristics of the Shocked Gas} \label{res-qua}
Since the cloud S2 shows no relevant feature in the IRC images and was not covered in the WIRC \HH{} image, we analyzed the cloud S1 only.
To quantify the infrared characteristics of the cloud S1, we measured its intensity in the IRC and WIRC images.
The  regions are outlined by the two concentric circles (see Fig.~\ref{fig-rgb}).
The inner circle is the source region and the surrounding annular shell is the background region.
To avoid any contamination during the measurement, possible point sources were excluded referring 2MASS point sources \citep{Skrutskie(2006)AJ_131_1163s}; they are indicated as white circles with a red slash on Figure \ref{fig-rgb}.
The White circles with black shadings on the IRC images (Fig.~\ref{fig-rgb}) are bright point sources masked out during the data reduction (cf.~$\S$~\ref{obs-irc}).
Also, the northern part of the IRC images was additionally excluded since the WIRC image does not fully cover this region.
The masked area is outlined by a white tetragon with a red slash.

Table \ref{tbl-result} lists the measured intensities together with the IRC colors, N4/S7 and S7/S11.
The IRC intensity is the strongest in S11, and decreases to shorter wavelength.
Comparing with Cloud N, the S7 and S11 intensities are greater by a factor of 2--3 in the cloud S1.
The colors are displayed as a point in Figure \ref{fig-ccd}.
The colors of the N2front in Cloud N are also displayed for comparison (cf.~section \ref{dis-ccd}).
The cloud S1 and N2front have similar colors.
The \HH{} intensity was also extinction-corrected, as in Cloud N.
We calculated the extinction factor to be $\sim0.82$ ($A_V=1.8$ mag), derived from the extinction curve of ``Milky Way, $R_V=3.1$'' \citep{Weingartner(2001)ApJ_548_296,Draine(2003)ARA&A_41_241} with the foreground hydrogen nuclei column density, $N$(H)=$N$(\Hi)+2$N$(\hh)$=(3.5\pm0.4)\times10^{21}$ \Ncm, towards the center of HB 21 \citep{Lee(2001)inproc}.

\section{Radiation Source of the Shock-Cloud Interaction Features Observed in the \akari{} IRC Bands} \label{irc-org}
To interpret the infrared intensities and colors (Table \ref{tbl-result}), we must identify the radiation source of the features we see in the shock-cloud interactions.
In Paper I, based on several arguments, we concluded that shocked \hh{} gas was the most probable explanation for the interaction features observed in the IRC S7, S11, and L15 bands.
In the similar manner, we here attribute the infrared features seen around the cloud S1 in the IRC N4, S7, and S11 bands to shocked \hh{} gas.

Firstly, the similarity between the features seen in the IRC images and the Palomar WIRC \HH{} image is also seen in the cloud S1 case.
Although the IRC images are rather diffuse, they definitely show three elongated clumpy features, which similarly locate around the cloud S1 as in the \HH{} image (Fig.~\ref{fig-rgb} and $\S$~\ref{res-mor}).
At least, this suggests that the features seen in the IRC bands arise partly from \hh{} emissions.
Secondly, the same observational and theoretical arguments for \hh{} emission, presented in Paper I, are valid over the IRC N4, S7, S11 bands ($\sim3-13$ \um): (1) only the \hh{} emission lines belong to the ``lines of S and \hh{}($J_{\textrm{\footnotesize{up}}}>2$)'' group can produce spatially similar features \citep{Neufeld(2007)ApJ_664_890}; (2) \hh{} lines are the dominant emission from shocked molecular gas whose physical parameters are similar to the cloud S1 (\vs$=20$ \kms, \nhh$=10^4$ \ncm; \citealt{Kaufman(1996)ApJ_456_611}).

We also considered other possible sources for the emission, presented in Paper I: fine structure ionic lines, thermal dust continuum, Polycyclic Aromatic Hydrocarbons (PAHs) bands, and synchrotron radiation.
Again, these do not seem likely either, as we discuss below.

\begin{itemize}
\item 
Within the wavelength coverage of the IRC N4, S7, S11 bands, there are three strong ionic lines, \Feii{} 5.34 \um, \Arii{} 6.99 \um, and \Neii{} 12.8 \um, which have been observed in the shocked regions of SNRs \citep[e.g.~][]{Arendt(1999)ApJ_521_234,Oliva(1999)A&A_343_943,Reach(2002)ApJ_564_302,Neufeld(2007)ApJ_664_890}.
The ionization potentials of these ions are 7.9 eV (\ionFeii), 15.8 eV (\ionArii), and 21.6 eV (\ionNeii), respectively.
From their case study on four SNRs, \cite{Neufeld(2007)ApJ_664_890} showed that the ions in shock-cloud interaction regions have \emph{two} distinctive spatial distributions according to their ionization potential, $>13.6$ eV and $<13.6$ eV.
Thus, in principle, the three ionic lines (\Feii, \Arii, and \Neii) can generate similar features in the IRC N4, S7, and S11 band images.
However, the ions observed in the shock-cloud interaction regions have a low correlation with \hh{} \citep{Neufeld(2007)ApJ_664_890}; indeed, such low correlations between \Feii{} and \hh{} have been frequently observed around SNRs \citep[e.g.~][]{Oliva(1999)A&A_343_943,Koo(2007)ApJ_657_308,Lee(2009)ApJ_691_1042}.
For the case of cloud S1, which shows a good correlation between the diffuse IRC and the \HH{} features (cf.~section \ref{res-mor}), it therefore seems that ionic lines from \Feii, \Arii, and \Neii{} are not responsible for the features we measured with AKARI.

\item
Thermal dust continuum is not likely responsible, either.
To produce a correlated features through the IRC N4, S7, and S11 bands, the dust temperature should be higher than 500 K.
In C-shocks, thought to be operating in the cloud S1, the dust temperature is below $\sim$50 K \citep{Draine(1983)ApJ_264_485}.
In addition, to heat up the dust over 500 K, there should be hot gas ($\gtrsim10^6$ K), well traced in X-rays, around the dust.
However, no significant hot gas exists around the cloud S1 \citep{Byun(2006)ApJ_637_283}.

\item
PAHs are another candidate since they are ubiquitous and have strong, broad band features at 3.3, 6.2, 7.7, 8.6, 11.2, 12.7, and 16.4 \um{} \citep{Tielens(2008)ARA&A_46_289}.
However, in the similar ways presented in Paper I, they are not likely to be the source for the diffuse features seen in the IRC bands.
PAHs are heated slowly and cool fast, thus it is hard to observe the shocked PAH emission above the background PAH emission \citep{Tielens(2008)ARA&A_46_289}; although one case was claimed to detect the shocked PAH emission \citep{Tappe(2006)ApJ_653_267}, hitherto, such emissions have not been observed in shocks \citep{vanDishoeck(2004)ARA&A_42_119,Tielens(2008)ARA&A_46_289}.

\item
Synchrotron radiation is unlikely, as well, since no obvious correlation between the \CO{} emission and the radio continuum was observed around the Cloud S \citep[section 4.2 of][]{Koo(2001)ApJ_552_175}.

\end{itemize}

\section{Comparison to Shock Models}
To interpret the infrared intensities and colors (Table \ref{tbl-result}), we must model the excitation of the features.
In Paper I, we concluded that shocked \hh{} gas was the most probable explanation, based on several arguments.
Among these, the clearest was the similarity of the diffuse features seen in the IRC and the WIRC \HH{} images.

Hence, here we now analyze the band intensities as emission lines from shocked \hh{} gas, which we calculate by applying several different shock models.
These models are the same as in Paper I, except for the non-stationary model.
This has been excluded since some bright \hh{} emission lines which fall into the IRC N4 band are not listed in the published models \citep[see Table 1 of][]{Flower(1999)MNRAS_308_271}.
Further descriptions of all these models are given in Paper I.

\subsection{C-Shock: Isothermal \hh{} Gas} \label{cshock-iso}
It is known that the shocked \hh{} gas behind a planar C-type shock can be approximated as an \emph{isothermal} and isobaric slab of gas \citep{Neufeld(2006)ApJ_649_816}, in view of the \hh{} excitation diagrams predicted for such shocks \citep[e.g.][]{Kaufman(1996)ApJ_456_611,Wilgenbus(2000)A&A_356_1010}.
Hence, we first calculate the expected IRC colors from the emission lines of \emph{isothermal} \hh{} gas.
Figure \ref{fig-ccd} displays the modeled IRC colors from isothermal \hh{} gas as open circles ($\circ$).
Their trajectory moves from the lower-left corner to the upper-right corner as the temperature increases (i.e. becomes increasingly 'blue').
This is explained because pure rotational lines of \hh, which are dominant below a few 1000 K in the IRC bands, have shorter wavelengths for higher upper-levels.
As \nhh{} increases, the populations are thermalized, approaching Local Thermodynamic Equilibrium (LTE).

As Figure \ref{fig-ccd} shows, isothermal \hh{} gas can not explain the observed IRC colors with any combination of \nhh{} and temperature.
The ortho-to-para ratio (OPR) was also varied from 0.5 to 5, since the OPR is expected to be different from 3.0 in the interstellar clouds \citep{Dalgarno(1973)ApL_14_77,Flower(1984)MNRAS_209_25,Lacy(1994)ApJ_428_L69} and even in shocked gas \citep{Timmermann(1998)ApJ_498_246,Wilgenbus(2000)A&A_356_1010}.
However, these variations are not able to reproduce the observed IRC colors (Fig.~\ref{fig-opr}).
The expected IRC colors at the \emph{same} temperature vary according to the adopted OPR; however, the \emph{locus} of the IRC colors do not differ much from the OPR=3.0 case, that is shown in Figure~\ref{fig-opr}.

A similar result was already found for the Cloud N, and it is consistent with the \hh{} level populations displaying an \emph{ankle-like curve} (see Fig.~\ref{fig-pop}).
The critical density of an \hh{} line transition increases as the energy level of the upper state increases \citep[cf.][]{LeBourlot(1999)MNRAS_305_802}; hence, isothermal \hh{} gas can only produce either a \emph{straight line} (LTE) or a \emph{knee-like curve} (non-LTE, see Fig.~1 in Paper I) in the population diagram, neither of which are observed.
Therefore, the cloud S1 also has an ankle-like \hh{} population.

This ankle-like population can be understood by the morphology of the diffuse \HH{} features.
These features are filamentary and surround the shocked CO cloud (cf. Fig.~\ref{fig-result} and \ref{fig-rgb}).
Hence, if they are generated by shocks propagating into the cloud S1, a range of shock velocities are expected.
Since the postshock temperature of C-shock is proportional to the shock velocity as \vs$^{1.35}$ \citep{Neufeld(2006)ApJ_649_816}, a moderate difference in \vs{} may result in a range of temperature in the shocked \hh{} gas.
However, this explanation may not be valid, because the same population was also found for the N2front of Cloud N, whose appearance is so planar that little difference in shock speed is expected (Paper I).

\subsection{C-Shock: Power-law Distribution of \hh{} Gas Temperature} \label{cshock-pow}
Figure \ref{fig-ccd} also displays the IRC colors calculated from the admixture model for \hh{} gas, as filled circles ($\bullet$).
As can be seen, it can reproduce the observed IRC color ratios with an appropriate combination of \nhh{} and $b$.
The derived parameters are listed in Table \ref{tbl-par}.
\NhhIRC{} is determined by scaling the modeled IRC intensity for the derived \nhh{} and $b$ to meet the observed IRC intensity.
The detail contributions of \hh{} line emission to IRC bands for these parameters are listed in Table~\ref{tbl-cont}.
The ``Weight'' column of Table~\ref{tbl-cont} lists the weighting factor for each line to the IRC band contribution.
For example, the S11 band intensity can be calculated as follows.
\begin{equation} \label{eq-cont}
\frac{I_{S11}(\textrm{\hh})}{\textrm{MJy sr$^{-1}$}}=\frac{0.610\,I[\textrm{\hh}\,S(2)]+0.921\,I[\textrm{\hh}\,S(3)]}{10^{-4}\,\textrm{\luerg}}
\end{equation}
As Table~\ref{tbl-cont} shows, the pure-rotational \hh{} emission lines are dominant in all IRC bands.
Since the N4 band, in contrast to the L15 band used for Cloud N, is used for the color-color diagram of the S1 cloud (Fig.~\ref{fig-ccd}), the contribution from several higher-level emission lines, S(7)--S(11), was also included, while that from the S(1) 17 \um{} emission line was not applied (cf.~Table 4 of Paper I).
We here note that the model parameters for N2front derived from the N4/S7 vs. S7/S11 color-color diagram (Fig.~\ref{fig-ccd}), $b\sim4$ and \nhh$\sim4\times10^4$ \ncm, are both \emph{larger} than those from the S7/S11 vs. S11/L15 diagram (Paper I), $b\sim3$ and \nhh$\sim2\times10^3$ \ncm.
We discuss this issue further in section \ref{dis-ccd}.

In addition, since \COv{} 4.6 $\mu m$ emission lines have been observed in shocked gas \citep[e.g.][]{Rosenthal(2000)A&A_356_705} and fall into the IRC N4 band coverage, we assessed their contribution to the band, for the power-law thermal admixture model with $b=4.0$ and $5.0$, referring the assessment of \cite{Neufeld(2008)ApJ_678_974}.
We used the CO vibrational energy state of \cite{Balakrishnan(2002)ApJ_568_443} and the CO vibrational transition rate of \cite{Chandra(1996)A&AS_117_557}.
We adopted the collisional rate coefficients for the excitation of CO vibrational transitions by H \citep{Balakrishnan(2002)ApJ_568_443} and by He \citep{Cecchi-Pestellini(2002)ApJ_571_1015}.
For excitation by \hh, we adopted the equations [7] and [8] of \cite{Thompson(1973)ApJ_181_1039} with the parameter $A$ of 68, the laboratory measurement of \cite{Millikan(1963)JChPh_39_3209}.
Unlike the excitation of \hh{} gas, we included H as a collisional partner for the CO vibrational excitation since H excites CO vibrational levels ($\upsilon>0$) \emph{more} efficiently than He and \hh.
H excites the \hh{} pure rotational levels ($\upsilon=0$) \emph{less} efficiently than He and \hh{} \citep{LeBourlot(1999)MNRAS_305_802}, hence including H as a collisional partner in the excitation of \hh{} gas makes negligible effects on the predicted IRC band intensity, where the pure rotational lines are dominant (cf. Table \ref{tbl-cont}).

Figure \ref{fig-cov} displays the results.
The fractional abundance of atomic hydrogen to molecular hydrogen, $N$(\Hi)/\Nhh, was varied as 0, 0.01, 0.1, and 1.0, while that of CO was fixed as $10^{-4}$.
As with the \hh{} vibrational states, those of CO also have higher collisional coefficients for a collision with atomic hydrogen than with He or \hh; hence, Figure \ref{fig-cov} shows a sensitive dependence on the ratio, $N$(\Hi)/\Nhh.
In the range of \nhh=$10^3-10^6$ \ncm, the contribution to the IRC N4 band is less than 0.1, hence negligible.
Furthermore, a robust simulation expects that \COv{} emission lines are much weaker than those of \hh{} in C-shocks of preshock \hh{} densities \nhh$=10^4-10^6$ \ncm{} and shock velocities \vs=20--40 \kms{} \citep{Kaufman(1996)ApJ_456_611}.

The derived parameters, except the power-law index $b$, are a little higher than those previously determined towards several SNRs, where interaction with nearby molecular clouds is occurring.
The density, \nhh=(3.9$_{-1.2}^{+2.1}$)$\times10^{4}$ \ncm, is a few times higher than the value, derived from Large Velocity Gradient (LVG) analysis of CO data for HB 21, of \nhh$=7.0\times10^3$ \ncm{} by \cite{Koo(2001)ApJ_552_175}.
The column density we derived, \NhhIRC=(2.8$_{-0.5}^{+0.2}$)$\times10^{21}$ \Ncm, is similarly higher than that derived towards shock-cloud interaction regions in four other SNRs (W 44, W 28, 3C 391, and IC 443), \Nhh{}$=(2.8-8.9)\times10^{20}$ \Ncm.
The latter values were determined from a two-temperature LTE fitting of pure-rotational \hh{} spectra with varying OPRs \citep{Neufeld(2007)ApJ_664_890}.
Our derived $b$-value of 4.2$_{-0.1}^{+0.1}$ falls into the middle of the range, 3.0--6.0, found by \cite{Neufeld(2008)ApJ_678_974}.
These authors found the IRAC color ratios to be well explained with this range of power-law index $b$, analyzing \spitzer{} IRAC observations towards the SNR IC 443.
From these three parameters derived---\nhh, \NhhIRC, and $b$---we also determined the model prediction for the \HH{} intensity, (1.5$_{-0.3}^{+0.5}$)$\times10^{-6}$ \luerg.
It is about a factor of four smaller than the observed value (see Table~\ref{tbl-result} and \ref{tbl-par}).
In contrast, for Cloud N the excess was found to be a factor of $17-33$ (Paper I).
We discuss these results further in $\S$\ref{discuss}.

Finally, we visualized the population state of the cloud S1, derived from the IRC color-color diagram, in Figure \ref{fig-pop} (left).
The pure-rotational levels which contribute to the IRC bands are designated with filled circles.
Also, the upper level of \HH{} emission line is designated with a filled triangle; its population derived from the observed \HH{} intensity, extinction corrected, is designated with a grey filled triangle with an error bar.
The diagram shows a severe deviation of $\upsilon>0$ levels from $\upsilon=0$ level, it thus seems that the two temperature LTE fitting, a model usually applied for shocked \hh{} gas \citep[e.g.][]{Rho(2001)ApJ_547_885,Giannini(2006)A&A_459_821}, does not properly describe the population state of the cloud S1, even with varying OPR.
This is caused by the low \nhh{} of the cloud S1, $\sim4\times10^4$ \ncm, which is much lower than the critical densities for ro-vibrational \hh{} lines, $\gtrsim10^8$ \ncm{} \citep{LeBourlot(1999)MNRAS_305_802}. 
We also note that this type of $\upsilon>0$ states population may not be easily recognizable with near-infrared \emph{ground} observations, since only the population for a few lowest-$J$ levels of the $\upsilon>0$ states can be deduced due to the atmospheric absorption \citep[e.g.][]{Burton(1989)inproca,Giannini(2006)A&A_459_821}.
Hence, for an exact derivation of the \hh{} level population, we must cover full range of $\sim2-30$ \um{} at once as in \cite{Rosenthal(2000)A&A_356_705}, and space observatories are ideal and mandatory in this sense.

\subsection{Partially Dissociative J-shock}
As Figure \ref{fig-other} shows, a partially dissociating jump-shock model does not reproduce the observed color of the cloud S1.
The observed color might appear to lie on a model extension to very high pressure, higher than $P=10^{11}$ \ncmK; however, this is implausible.
From their CO observations, \cite{Koo(2001)ApJ_552_175} derived \nhh$=7.0\times10^3$ \ncm{} and \vs$\lesssim20$\kms{} for the cloud S1.
These give a postshock pressure of $\sim10^8$ \ncmK, which is more than $10^3$ times lower than the above limit.
A pressure enhancement can occur for the collision between molecular clumps and radiative shells of a remnant \citep[e.g.][]{Moorhouse(1991)MNRAS_253_662,Chevalier(1999)ApJ_511_798}; however, it is only about a factor of 20.
Insufficient cooling time cannot solve the disagreement between the observed and modeled colors, either.
If the postshock \hh{} had not cooled as low as a few hundred K, then the modeled IRC colors move towards the upper-right direction in the color-color diagram (Fig.~\ref{fig-ccd}) to bluer colors.
This was not observed.
Overall, a partially dissociative J-shock does not seem to be a suitable model to explain the observed IRC colors.

\section{Discussion} \label{discuss}
The observed color ratios were only reproduced by the thermal admixture model, as was the case for Cloud N (Paper I).
Hence, we here discuss the derived parameters from this model, based on pictures for the shock-cloud interactions, as proposed in \cite{Neufeld(2008)ApJ_678_974} and in Paper I.
We also note here that we assumed the OPR=3.0 since no OPR information is available for the cloud S1; hence, the derived parameters can be changed according to the \emph{adopted} OPR value.

\subsection{Nature of Molecular Shocks Seen in the Infrared} \label{dis-nat}
Diffuse infrared features surround the shocked CO cloud S1 (see Fig.~\ref{fig-result} and \ref{fig-rgb}).
For instance, the \HH{} image shows several filamentary features, as if excited by shocks propagating into the cloud core.
These may represent distinctive planar shocks, each with different speed.
From the previous observations toward SNR molecular shock regions, the \hh{} level population diagram has been shown to have an ankle-like curve (cf.~Fig.~\ref{fig-pop}).
For planar C-shocks to explain such populations, it generally requires two components, whose shock velocities are $\sim10$ and $\sim30-50$ \kms{}, with comparable amounts of \Nhh{} \citep[cf.~][]{Hewitt(2009)ApJ_694_1266}.
Hence, the filamentary features seen in the \HH{} image may originate from such a mixture of planar C-shocks.
However, the possibility that the \emph{individual} filamentary feature bears the ankle-like population still remains, since such a property was seen in a very planar filamentary feature of Cloud N (N2front).

\cite{Neufeld(2008)ApJ_678_974} showed that the $b$ values they obtained, $\sim3.0-6.0$, can be explained by paraboloidal bow shocks, which are geometrical summations of planar C-shocks (see Fig.~\ref{fig-pic}).
In their picture, a paraboloidal bow shock, where \hh{} survives the shock (i.e. T$\lesssim4,000$ K), has a power-law index $b\sim3.8$.
If some slower bow-shocks which do not reach 4000 K are then spatially averaged together, a value for $b$ of $\gtrsim3.8$ is generated.

The value $b$ for the cloud S1 was determined to be 4.2$_{-0.1}^{+0.1}$, which falls into the range derived for bow shocks, $b\gtrsim3.8$.
However, as discussed in Paper I, bow shocks should have been observed in the \HH{} image, if any, since the expected shock width for a planar C-shock propagating into preshock gas of \nhh{} $\sim10^4$ \ncm{} is $\sim10^{16}$ cm \citep{Draine(1983)ApJ_264_485}, comparable to the spatial resolution in the image, $\sim1.1''$ (cf.~$\S$~\ref{obs-wirc}) $\sim(1.3-2.8)\times10^{16}$ cm for the distance of $\sim0.8-1.7$ kpc \citep{Leahy(1987)MNRAS_228_907,Tatematsu(1990)A&A_237_189,Byun(2006)ApJ_637_283}.
This absence of bow shock features can be explained by the viewing angle.
If we consider the circular and filamentary appearance of the diffuse \hh{} features around the cloud S1, it may be possible that a single paraboloidal bow-shock is being viewed along its symmetry axis, producing the circular feature seen in Figure \ref{fig-rgb}.
In this case, our result fit with the bow shock picture of \cite{Neufeld(2008)ApJ_678_974}, when seen face-on.

This bow shock picture has a difficulty in achieving a steady state for the shock, however.
It assumes a steady state planar C-shock at every point of the bow.
Through the C-type shock, the \nhh{} can be increased up to a factor of ten \citep[e.g.][]{Timmermann(1998)ApJ_498_246,Wilgenbus(2000)A&A_356_1010}.
Thus, \nhh{} at the upstream of the bow would be $\sim$ [\nhh{} at downstream]$\times0.1\sim[4\times10^4]\times0.1\sim4\times10^3$ \ncm{} (cf.~Table~\ref{tbl-par} and Fig.~\ref{fig-pic}).
Also, the preshock gas velocity into the shock is known to be $\sim20$ \kms{} from CO observations \citep{Koo(2001)ApJ_552_175}.
For these preshock density and shock velocity, the time required to achieve a steady shock is known to be $\sim10^4$ yr from the study on the early stage of shock generation \citep{Flower(1999)MNRAS_308_271}.
This time seems to be long for the bow shock around the cloud S1 to be in a steady state, considering the estimated age of HB 21, $\sim5000-7000$ yr \citep{Lazendic(2006)ApJ_647_350,Byun(2006)ApJ_637_283}, together with the location of the cloud S1 near the edge of the remnant (Fig.~\ref{fig-obs}); we here note that the remnant may be older than 5000-7000 yr, estimated at the distance of 0.8 kpc, since the distance is uncertain, $\sim0.8-1.7$ kpc \citep{Leahy(1987)MNRAS_228_907,Tatematsu(1990)A&A_237_189,Byun(2006)ApJ_637_283}.
	
In Paper I, we conjectured that a shocked clumpy interstellar medium (ISM) exists (cf.~Fig.~\ref{fig-pic}), based on the similar $b$ values of the N2front and N2clump regions and on the cyclodial (cuspy) feature seen in the N2clump region, together with numerical simulations \citep{Nakamura(2006)ApJS_164_477,Shin(2008)ApJ_680_336}.
If this picture also holds for the S1 cloud, the shocked clump must be unresolved in the WIRC \HH{} image, since the \hh{} features around this cloud do not show any cycloidal features.
However, even though this is the case, the absence of the wriggle expected for shock fronts propagating a clumpy ISM \citep[e.g.][]{Patnaude(2005)ApJ_633_240} still remains as an issue (cf.~Fig.~\ref{fig-rgb})---the wriggle is generated by shocks propagating further through a lower density medium, and vice versa.
This wriggle can be unresolved in the \HH{} image if its scale is small enough ($\lesssim10^{16}$ cm).
However, it is uncertain whether the cycloidal feature would be maintained under such a small scale.

As noted in section \ref{res-qua}, the N4/S7 vs. S7/S11 colors of the cloud S1 and N2front are similar (cf.~Fig.~\ref{fig-ccd}).
This is intriguing considering that they are physically unrelated.
Their morphologies seen in \HH{} images are also similar, i.e.~filamentary, although their sizes show a few factors of difference (cf.~ Paper I and Fig.~\ref{fig-rgb}).
These similarities suggest that the cloud S1 and N2front share similar shock conditions.
The interstellar ultraviolet radiation field may contribute to this similarity; however, a more robust study is required.

\subsection{\HH{} intensity} \label{dis-h2s1}
We estimated the \HH{} intensity of the cloud S1 for the derived model parameters---\nhh, $b$, and \NhhIRC---from the mid-infrared IRC colors.
It is about four times weaker than the observed intensity (Table \ref{tbl-result} and \ref{tbl-par}).
This discrepancy is less severe than the Cloud N case, which shows a factor of 17--33 difference (Paper I).
However, the amount of excited gas, $N(\textrm{\hh};v=1,J=3)$, required to compensate for the difference is $\sim10^{14}$ \Ncm{} in both cases (cf.~Fig.~\ref{fig-pop}).

In Paper I, we discussed two possible reasons for the discrepancy.
Firstly, the existence of additional \hh{} gas, whose temperature and density are both high, but whose column density is low enough to have negligible effect on the mid-infrared line intensities.
For example, to compensate for a deficiency of $N(v=1,J=3)\sim10^{14}$ \Ncm, we need an additional amount of \hh{} gas of \Nhh{} $\sim10^{16}$ \Ncm{} in LTE with $T\sim2000$ K.
A compact, unresolved shocked cloud is a candidate for such additional \hh{} gas.

The second explanation given was the omission of collisions with hydrogen atoms, which are effective in exciting the vibrational states of \hh{} \citep[cf.~][]{Neufeld(2008)ApJ_678_974}.
The cross section for excitation of \hh{} by H is several orders of magnitude greater for rovibrational transitions than it is for pure rotational transitions \citep[see Table~1 and Figure~1 in][]{LeBourlot(1999)MNRAS_305_802}.
Hence, with only a small fraction of H, \nh{}/\nhh{}$\sim0.025$, the rovibrational transition can be dominated by collisions with H, rather than with \hh, in the temperature range 300--4000 K.
Indeed, such a fraction of atomic gas is expected in interstellar clouds with \nhh{} $\gtrsim10^3$ \ncm{} \cite[see Table~1 and Figure~1 in][]{Snow(2006)ARA&A_44_367}, as well as in theoretical models for shock waves that are fast enough to produce \hh{} at temperatures of a few thousand K \citep[e.g.][]{Wilgenbus(2000)A&A_356_1010}.

\subsection{Limitation of the Thermal Admixture Model}

\subsubsection{\hh{} Column Density \Nhh}\label{dis-col}
In section \ref{cshock-pow}, we mentioned that the column density \NhhIRC{} of the cloud S1 is a few times higher than those of other SNRs (W 44, W 28, 3C 391, and IC 443).
The former is \NhhIRC=(2.8$_{-0.5}^{+0.2}$)$\times10^{21}$ \Ncm, while the latter are \Nhh{}$=(2.8-8.9)\times10^{20}$ \Ncm{} \citep{Neufeld(2007)ApJ_664_890}.
This seems unreasonable considering that the \nhh{} of the cloud S1, (3.9$_{-1.2}^{+2.1}$)$\times10^{4}$ \ncm, is lower than that of IC 443, $\sim10^7$ \ncm{} \citep{Neufeld(2008)ApJ_678_974}. 
However, there is one important point to note before the comparison: those \hh{} column densities are \emph{expectations} estimated using different methods.

The total column density of \hh{} gas at a few 100 K is mainly determined by the values of $\upsilon=0$ $J=0,1$ levels, since other levels have much lower populations (cf.~Fig.~\ref{fig-pop}).
However, we cannot obtain the column densities of these levels directly from emission lines, because no transition to a lower state is allowed; therefore, we must estimate the column densities of $J=0,1$ levels from the observed column densities of $J>1$ levels.
The different methods used for this estimation causes the discrepancy between the cloud S1 and IC 443, mentioned above, as we discuss below.

The \NhhIRC{} of the cloud S1, (2.8$_{-0.5}^{+0.2}$)$\times10^{21}$ \Ncm, is estimated by applying the thermal admixture model to the observed IRC intensities, while the \Nhh{} of IC 443, $\sim5.0\times10^{20}$ \Ncm, was estimated with two temperature LTE fitting with varying the OPR \citep{Neufeld(2007)ApJ_664_890}.
If we estimate the \Nhh{} of the cloud S1 using the same two-temperature LTE fitting, we would obtain a lower values.
Figure \ref{fig-pop} (top-right) displays the result of such fitting.
The fitting was applied only to the \hh{} levels whose emission lines contribute to the IRC bands (cf. Table \ref{tbl-cont}), and returned a column density of $\sim7.3\times10^{19}$ \Ncm.
This is about seven times less than IC 443, and now does not cause the discrepancy mentioned above.

Also, this column density is about \emph{40 times smaller} than the estimation using the thermal admixture model.
This difference stems from the fact that the thermal admixture model does not estimate the column densities of $J=0,1$ with a linear extrapolation in the population diagram as the LTE fitting does (cf.~Fig.~\ref{fig-pop}); it estimates the column densities with a curved population, defined by the equation $dN\sim T^{-b}dT$ (100 K $\le T \le$ 4000 K).
Such a difference is smaller when longer wavelength IRC bands are used for the column density estimation.
Figure \ref{fig-pop} (bottom-right) shows the results of the two temperature LTE fitting for the population of the N2front, determined using the IRC S7, S11, and L15 bands (Paper I); the result shows a column density of $\sim7.5\times10^{19}$ \Ncm, which is about \emph{3.3 times smaller} than the estimation using the thermal admixture model, $\sim2.5\times10^{20}$ \Ncm{} (Paper I).
This trend is reasonable since the \emph{longer wavelength} IRC bands determine the level populations of \emph{lower-$J$} states in $\upsilon=0$, which results in a less severe extrapolation for the column densities, $N(J=0,1)$.

\subsubsection{\hh{} Density \nhh{} and power-law index $b$} \label{dis-ccd}
To compare the N4/S7 and S7/S11 colors of the cloud S1 with those of N2front in Cloud N (Fig.~\ref{fig-ccd}), we further determined the N4 intensity of N2front, which was not measured in Paper I because of strong point source contamination.
In order to remove this contamination, we additionally masked point-source dominated areas seen in the N4 band.
Table \ref{tbl-n2f} shows the measured intensities.
Since some areas are additionally masked unlike Paper I, the intensities of S7, S11, and L15 are a little different from those listed in Paper I; however, the colors S7/S11 and S11/L15 are almost unchanged (cf.~Fig.~7 in Paper I and Fig.~\ref{fig-n2f}). 
Therefore, we think the point-source masking was done properly.

Figure \ref{fig-n2f} shows that the model parameters, $b$ and \nhh, obtained for N2front depends on which IRC bands are used for the color-color diagram.
The diagram of shorter wavelength bands (N4, S7, S11; Fig.~\ref{fig-n2f} leftmost) returns \emph{larger-b} and \emph{larger-\nhh} values than the diagram of longer wavelength bands (S7, S11, L15; Fig.~\ref{fig-n2f} rightmost), and the diagram of N4/S7 vs. S11/L15 (Fig.~\ref{fig-n2f} middle) returns the middle values of the former two diagrams'.
This inconsistency may be caused by the intrinsic property of shocked \hh{} gas.
In other words, the whole level population of shocked \hh{} gas may not be fully described by the power-law thermal admixture model with \emph{only one set} of $b$ and \nhh.

To check this possibility, in Figure \ref{fig-n2f}, we overplotted the IRC colors of Orion Molecular Cloud-1 (OMC-1), where extensive emission lines of shocked \hh{} gas were observed over $2.5-30$ \um{} \citep{Rosenthal(2000)A&A_356_705}.
The OMC-1 emissions were adjusted to experience the same extinction with HB 21, to be placed in the model grid for N2front.
Interestingly, OMC-1 also shows the same trend for the model parameters, $b$ and \nhh, as the case for N2front.
This suggests that the variations of $b$ and \nhh{} are needed for the thermal admixture model to describe the whole level population of shocked \hh{} gas.
No significant variation of $b$ was seen in the SNR IC 443, where the thermal admixture model was applied first \citep{Neufeld(2008)ApJ_678_974}.
It may be caused by the narrow wavelength coverage of the bands they used (\spitzer{} IRAC; $\sim3-8$ \um), which missed the longer wavelength information we used.
We here note that the model parameters---$b$ and \nhh---must be obtained from the \emph{same} band images for their comparison between different shocked regions, since the parameters are likely dependent on the wavelength.

\section{Conclusion}
We have observed a shock-cloud interaction region in the SNR HB 21 at near- and mid-infrared wavelengths, with the WIRC at the Palomar telescope and the IRC aboard the \akari{} satellite.
The IRC N4, S7, and S11 band images and the WIRC \HH{} image reveal similar diffuse features, which surround the shocked CO cloud S1.
However, there are no infrared diffuse features seen around another shocked CO cloud S2.
Lack of shocked \hh{} gas may cause this absence, but why it happens only for the cloud S2 is uncertain.

We found that the IRC colors of the cloud S1 are well explained by an admixture model of \hh{} gas temperatures, whose infinitesimal column density varies as $dN\sim T^{-b}dT$.
Three physical parameters---\nhh, $b$, and \NhhIRC---were derived from this thermal admixture model (cf.~Table~\ref{tbl-par}).
These can be understood with multiple planar C-shocks whose velocities are different.
Alternatively, the derived $b$ value ($\sim4.2$) can be understood through a bow shock picture, if we are looking at a single bow shock along the symmetry axis.
However, this picture has a difficulty in achieving a steady state.
A shocked clumpy ISM picture, conjectured in Paper I, remains as a possible explanation, but the absence of the wriggle, expected for shock fronts propagating a clumpy medium, in the filamentary features seen in \HH{} image remains as an issue.
The model parameters, $b$ and \nhh, obtained for the cloud S1 and N2front (cf.~Fig.~\ref{fig-ccd}) are very similar, which means that these clouds share similar shock conditions.

We also compared the observed \HH{} intensity to that predicted from the power-law admixture model.
It is about four times greater.
This excess might be caused by either an additional component of hot, dense \hh{} gas (which has low total column density), or by the omission of collisions with hydrogen atoms in the power-law admixture model (which results in an under-prediction of the near-IR line intensity).

The limitation of the thermal admixture model is explored with respect to the derived model parameters.
The \Nhh{} estimation of the model shows a smaller difference with those of two temperature LTE fitting, when longer wavelength IRC bands are used for the determination of model parameters.
Investigating the infrared colors of N2front and OMC-1 in the four IRC bands (N4, S7, S11, and L15), we found that the thermal admixture model cannot describe the whole \hh{} level population with only one set of $b$ and \nhh; the shorter wavelength bands returns higher-$b$ and higher-\nhh.
This tells we must use the same bands in determining the model parameters, for the comparisons of the shocked \hh{} gas' properties.


\section*{Acknowledgments}
This work is based on observations with \akari, a JAXA project with the participation of ESA. 
The authors thank all the members of the \akari{} project.
Also, the authors thank the referee for all the comments which make this paper clearer.
This work was supported by the Korea Science and Engineering Foundation (R01-2007-000-20336-0) and also through the KOSEF-NSERC Cooperative Program (F01-2007-000-10048-0).
This research has made use of SAOImage DS9, developed by Smithsonian Astrophysical Observatory \citep{Joye(2003)inproc}.

\bibliographystyle{G:/Work/Publication/bibtex/astronat/asr/elsart-harv}
\bibliography{G:/Work/Publication/bibtex/jhshinn}

\begin{thebibliography}{57}
\expandafter\ifx\csname natexlab\endcsname\relax\def\natexlab#1{#1}\fi
\expandafter\ifx\csname url\endcsname\relax
  \def\url#1{\texttt{#1}}\fi
\expandafter\ifx\csname urlprefix\endcsname\relax\def\urlprefix{URL }\fi

\bibitem[{{Arendt} et~al.(1999){Arendt}, {Dwek}, and
  {Moseley}}]{Arendt(1999)ApJ_521_234}
{Arendt}, R.~G., {Dwek}, E., {Moseley}, S.~H., Aug. 1999. {Newly Synthesized
  Elements and Pristine Dust in the Cassiopeia A Supernova Remnant}. ApJ 521,
  234--245.

\bibitem[{{Balakrishnan} et~al.(2002){Balakrishnan}, {Yan}, and
  {Dalgarno}}]{Balakrishnan(2002)ApJ_568_443}
{Balakrishnan}, N., {Yan}, M., {Dalgarno}, A., Mar. 2002. {Quantum-Mechanical
  Study of Rotational and Vibrational Transitions in CO Induced by H Atoms}.
  ApJ 568, 443--447.

\bibitem[{{Brand} et~al.(1988){Brand}, {Moorhouse}, {Burton}, {Geballe},
  {Bird}, and {Wade}}]{Brand(1988)ApJ_334_L103}
{Brand}, P.~W.~J.~L., {Moorhouse}, A., {Burton}, M.~G., {Geballe}, T.~R.,
  {Bird}, M., {Wade}, R., Nov. 1988. {Ratios of molecular hydrogen line
  intensities in shocked gas - Evidence for cooling zones}. ApJ 334,
  L103--L106.

\bibitem[{{Burton} et~al.(1989){Burton}, {Brand}, {Moorhouse}, and
  {Geballe}}]{Burton(1989)inproca}
{Burton}, M., {Brand}, P., {Moorhouse}, A., {Geballe}, T., Sep. 1989.
  {High-excitation lines of molecular hydrogen: A discriminant between shock
  models}. In: {B{\"o}hm-Vitense}, E. (Ed.), Infrared Spectroscopy in
  Astronomy. Vol. 290 of ESA Special Publication. Paris, France : European
  Space Agency, p. 281.

\bibitem[{{Byun} et~al.(2006){Byun}, {Koo}, {Tatematsu}, and
  {Sunada}}]{Byun(2006)ApJ_637_283}
{Byun}, D.-Y., {Koo}, B.-C., {Tatematsu}, K., {Sunada}, K., Jan. 2006.
  {Interaction between the Supernova Remnant HB 21 and Molecular Clouds}. ApJ
  637, 283--295.

\bibitem[{{Cecchi-Pestellini} et~al.(2002){Cecchi-Pestellini}, {Bodo},
  {Balakrishnan}, and {Dalgarno}}]{Cecchi-Pestellini(2002)ApJ_571_1015}
{Cecchi-Pestellini}, C., {Bodo}, E., {Balakrishnan}, N., {Dalgarno}, A., Jun.
  2002. {Rotational and Vibrational Excitation of CO Molecules by Collisions
  with $^{4}$He Atoms}. ApJ 571, 1015--1020.

\bibitem[{{Chandra} et~al.(1996){Chandra}, {Maheshwari}, and
  {Sharma}}]{Chandra(1996)A&AS_117_557}
{Chandra}, S., {Maheshwari}, V.~U., {Sharma}, A.~K., Jun. 1996. {Einstein
  A-coefficients for vib-rotational transitions in CO.} A\&AS 117, 557--559.

\bibitem[{{Chevalier}(1999)}]{Chevalier(1999)ApJ_511_798}
{Chevalier}, R.~A., Feb. 1999. {Supernova Remnants in Molecular Clouds}. ApJ
  511, 798--811.

\bibitem[{{Dalgarno} et~al.(1973){Dalgarno}, {Black}, and
  {Weisheit}}]{Dalgarno(1973)ApL_14_77}
{Dalgarno}, A., {Black}, J.~H., {Weisheit}, J.~C., 1973. {Ortho-Para
  Transitions in H$_2$ and the Fractionation of HD}. ApL 14, 77.

\bibitem[{{Draine}(2003)}]{Draine(2003)ARA&A_41_241}
{Draine}, B.~T., 2003. {Interstellar Dust Grains}. ARA\&A 41, 241--289.

\bibitem[{{Draine} et~al.(1983){Draine}, {Roberge}, and
  {Dalgarno}}]{Draine(1983)ApJ_264_485}
{Draine}, B.~T., {Roberge}, W.~G., {Dalgarno}, A., Jan. 1983.
  {Magnetohydrodynamic shock waves in molecular clouds}. ApJ 264, 485--507.

\bibitem[{{Erkes} and {Dickel}(1969)}]{Erkes(1969)AJ_74_840}
{Erkes}, J.~W., {Dickel}, J.~R., Aug. 1969. {Radio Observations of the
  Supernova Remnant HB 21}. AJ 74, 840.

\bibitem[{{Flower} and {Pineau des
  For{\^e}ts}(1999)}]{Flower(1999)MNRAS_308_271}
{Flower}, D.~R., {Pineau des For{\^e}ts}, G., Sep. 1999. {H\_2 emission from
  shocks in molecular outflows: the significance of departures from a
  stationary state}. MNRAS 308, 271--280.

\bibitem[{{Flower} and {Watt}(1984)}]{Flower(1984)MNRAS_209_25}
{Flower}, D.~R., {Watt}, G.~D., Jul. 1984. {On the ortho-H2/para-H2 ratio in
  molecular clouds}. MNRAS 209, 25--31.

\bibitem[{{Giannini} et~al.(2006){Giannini}, {McCoey}, {Nisini}, {Cabrit},
  {Caratti o Garatti}, {Calzoletti}, and {Flower}}]{Giannini(2006)A&A_459_821}
{Giannini}, T., {McCoey}, C., {Nisini}, B., {Cabrit}, S., {Caratti o Garatti},
  A., {Calzoletti}, L., {Flower}, D.~R., Dec. 2006. {Molecular line emission in
  HH54: a coherent view from near to far infrared}. A\&A 459, 821--835.

\bibitem[{{Hewitt} et~al.(2009){Hewitt}, {Rho}, {Andersen}, and
  {Reach}}]{Hewitt(2009)ApJ_694_1266}
{Hewitt}, J.~W., {Rho}, J., {Andersen}, M., {Reach}, W.~T., Apr. 2009. {Spitzer
  Observations of Molecular Hydrogen in Interacting Supernova Remnants}. ApJ
  694, 1266--1280.

\bibitem[{{Huang} and {Thaddeus}(1986)}]{Huang(1986)ApJ_309_804}
{Huang}, Y.-L., {Thaddeus}, P., Oct. 1986. {Molecular clouds and supernova
  remnants in the outer galaxy}. ApJ 309, 804--821.

\bibitem[{{Joye} and {Mandel}(2003)}]{Joye(2003)inproc}
{Joye}, W.~A., {Mandel}, E., 2003. {New Features of SAOImage DS9}. In: {Payne},
  H.~E., {Jedrzejewski}, R.~I., {Hook}, R.~N. (Eds.), Astronomical Data
  Analysis Software and Systems XII. Vol. 295 of Astronomical Society of the
  Pacific Conference Series. p. 489.

\bibitem[{{Kaufman} and {Neufeld}(1996)}]{Kaufman(1996)ApJ_456_611}
{Kaufman}, M.~J., {Neufeld}, D.~A., Jan. 1996. {Far-Infrared Water Emission
  from Magnetohydrodynamic Shock Waves}. ApJ 456, 611.

\bibitem[{{Koo} et~al.(2007){Koo}, {Moon}, {Lee}, {Lee}, and
  {Matthews}}]{Koo(2007)ApJ_657_308}
{Koo}, B.-C., {Moon}, D.-S., {Lee}, H.-G., {Lee}, J.-J., {Matthews}, K., Mar.
  2007. {[Fe II] and H$_2$ Filaments in the Supernova Remnant G11.2-0.3:
  Supernova Ejecta and Presupernova Circumstellar Wind}. ApJ 657, 308--317.

\bibitem[{{Koo} et~al.(2001){Koo}, {Rho}, {Reach}, {Jung}, and
  {Mangum}}]{Koo(2001)ApJ_552_175}
{Koo}, B.-C., {Rho}, J., {Reach}, W.~T., {Jung}, J., {Mangum}, J.~G., May 2001.
  {Shocked Molecular Gas in the Supernova Remnant HB 21}. ApJ 552, 175--188.

\bibitem[{{Lacy} et~al.(1994){Lacy}, {Knacke}, {Geballe}, and
  {Tokunaga}}]{Lacy(1994)ApJ_428_L69}
{Lacy}, J.~H., {Knacke}, R., {Geballe}, T.~R., {Tokunaga}, A.~T., Jun. 1994.
  {Detection of absorption by H$_2$ in molecular clouds: A direct measurement
  of the H$_2$:CO ratio}. ApJ 428, L69--L72.

\bibitem[{{Lazendic} and {Slane}(2006)}]{Lazendic(2006)ApJ_647_350}
{Lazendic}, J.~S., {Slane}, P.~O., Aug. 2006. {Enhanced Abundances in Three
  Large-Diameter Mixed-Morphology Supernova Remnants}. ApJ 647, 350--366.

\bibitem[{{Le Bourlot} et~al.(1999){Le Bourlot}, {Pineau des For{\^e}ts}, and
  {Flower}}]{LeBourlot(1999)MNRAS_305_802}
{Le Bourlot}, J., {Pineau des For{\^e}ts}, G., {Flower}, D.~R., May 1999. {The
  cooling of astrophysical media by H\_2}. MNRAS 305, 802--810.

\bibitem[{{Leahy}(1987)}]{Leahy(1987)MNRAS_228_907}
{Leahy}, D.~A., Oct. 1987. {Einstein X-ray observations of the supernova
  remnant HB21}. MNRAS 228, 907--913.

\bibitem[{{Leahy} and {Aschenbach}(1996)}]{Leahy(1996)A&A_315_260}
{Leahy}, D.~A., {Aschenbach}, B., Nov. 1996. {ROSAT X-ray observations of the
  supernova remnant HB 21.} A\&A 315, 260--264.

\bibitem[{{Lee} et~al.(2009){Lee}, {Moon}, {Koo}, {Lee}, and
  {Matthews}}]{Lee(2009)ApJ_691_1042}
{Lee}, H.-G., {Moon}, D.-S., {Koo}, B.-C., {Lee}, J.-J., {Matthews}, K., Feb.
  2009. {Near-Infrared [Fe II] and H$_2$ Line Observations of the Supernova
  Remnant 3C 396: Probing the Presupernova Circumstellar Materials}. ApJ 691,
  1042--1049.

\bibitem[{{Lee} et~al.(2001){Lee}, {Rho}, {Koo}, {Petre}, and
  {Decourchelle}}]{Lee(2001)inproc}
{Lee}, H.-G., {Rho}, J., {Koo}, B.-C., {Petre}, R., {Decourchelle}, A., May
  2001. {ASCA/ROSAT observations of the SNR HB 21}. In: Bulletin of the
  American Astronomical Society. Vol.~33 of Bulletin of the American
  Astronomical Society. p. 839.

\bibitem[{{Lorente} et~al.(2007){Lorente}, {Onaka}, {Ita}, {Ohyama}, and
  {Pearson}}]{Lorente(2007)man}
{Lorente}, R., {Onaka}, T., {Ita}, Y., {Ohyama}, Y., {Pearson}, C., 2007.
  \textit{AKARI} IRC Data User Manual Version 1.2.

\bibitem[{{Maloney} et~al.(1996){Maloney}, {Hollenbach}, and
  {Tielens}}]{Maloney(1996)ApJ_466_561}
{Maloney}, P.~R., {Hollenbach}, D.~J., {Tielens}, A.~G.~G.~M., Jul. 1996.
  {X-Ray--irradiated Molecular Gas. I. Physical Processes and General Results}.
  ApJ 466, 561.

\bibitem[{{Millikan} and {White}(1963)}]{Millikan(1963)JChPh_39_3209}
{Millikan}, R.~C., {White}, D.~R., Dec. 1963. {Systematics of Vibrational
  Relaxation}. JChPh 39, 3209--3213.

\bibitem[{{Moorhouse} et~al.(1991){Moorhouse}, {Brand}, {Geballe}, and
  {Burton}}]{Moorhouse(1991)MNRAS_253_662}
{Moorhouse}, A., {Brand}, P.~W.~J.~L., {Geballe}, T.~R., {Burton}, M.~G., Dec.
  1991. {Surprisingly high-pressure shocks in the supernova remnant IC 443}.
  MNRAS 253, 662--668.

\bibitem[{{Murakami et al.}(2007)}]{Murakami(2007)PASJ_59_S369s}
{Murakami et al.}, Aug. 2007. {The Infrared Astronomical Mission AKARI}. PASJ
  59, S369.

\bibitem[{{Nakamura} et~al.(2006){Nakamura}, {McKee}, {Klein}, and
  {Fisher}}]{Nakamura(2006)ApJS_164_477}
{Nakamura}, F., {McKee}, C.~F., {Klein}, R.~I., {Fisher}, R.~T., Jun. 2006. {On
  the Hydrodynamic Interaction of Shock Waves with Interstellar Clouds. II. The
  Effect of Smooth Cloud Boundaries on Cloud Destruction and Cloud Turbulence}.
  ApJS 164, 477--505.

\bibitem[{{Neufeld} et~al.(2007){Neufeld}, {Hollenbach}, {Kaufman}, {Snell},
  {Melnick}, {Bergin}, and {Sonnentrucker}}]{Neufeld(2007)ApJ_664_890}
{Neufeld}, D.~A., {Hollenbach}, D.~J., {Kaufman}, M.~J., {Snell}, R.~L.,
  {Melnick}, G.~J., {Bergin}, E.~A., {Sonnentrucker}, P., Aug. 2007. {Spitzer
  Spectral Line Mapping of Supernova Remnants. I. Basic Data and Principal
  Component Analysis}. ApJ 664, 890--908.

\bibitem[{{Neufeld} et~al.(2006){Neufeld}, {Melnick}, {Sonnentrucker},
  {Bergin}, {Green}, {Kim}, {Watson}, {Forrest}, and
  {Pipher}}]{Neufeld(2006)ApJ_649_816}
{Neufeld}, D.~A., {Melnick}, G.~J., {Sonnentrucker}, P., {Bergin}, E.~A.,
  {Green}, J.~D., {Kim}, K.~H., {Watson}, D.~M., {Forrest}, W.~J., {Pipher},
  J.~L., Oct. 2006. {Spitzer Observations of HH 54 and HH 7-11: Mapping the
  H$_{2}$ Ortho-to-Para Ratio in Shocked Molecular Gas}. ApJ 649, 816--835.

\bibitem[{{Neufeld} and {Yuan}(2008)}]{Neufeld(2008)ApJ_678_974}
{Neufeld}, D.~A., {Yuan}, Y., May 2008. {Mapping Warm Molecular Hydrogen with
  the Spitzer Infrared Array Camera (IRAC)}. ApJ 678, 974--984.

\bibitem[{{Oliva} et~al.(1999){Oliva}, {Moorwood}, {Drapatz}, {Lutz}, and
  {Sturm}}]{Oliva(1999)A&A_343_943}
{Oliva}, E., {Moorwood}, A.~F.~M., {Drapatz}, S., {Lutz}, D., {Sturm}, E., Mar.
  1999. {Infrared spectroscopy of young supernova remnants heavily interacting
  with the interstellar medium. I. Ionized species in RCW 103}. A\&A 343,
  943--952.

\bibitem[{{Onaka et al.}(2007)}]{Onaka(2007)PASJ_59_S401s}
{Onaka et al.}, May 2007. {The Infrared Camera (IRC) for AKARI - Design and
  Imaging Performance}. PASJ 59, S401.

\bibitem[{{Patnaude} and {Fesen}(2005)}]{Patnaude(2005)ApJ_633_240}
{Patnaude}, D.~J., {Fesen}, R.~A., Nov. 2005. {Model Simulations of a
  Shock-Cloud Interaction in the Cygnus Loop}. ApJ 633, 240--247.

\bibitem[{{Reach} et~al.(2002){Reach}, {Rho}, {Jarrett}, and
  {Lagage}}]{Reach(2002)ApJ_564_302}
{Reach}, W.~T., {Rho}, J., {Jarrett}, T.~H., {Lagage}, P.-O., Jan. 2002.
  {Molecular and Ionic Shocks in the Supernova Remnant 3C 391}. ApJ 564,
  302--316.

\bibitem[{{Rho} et~al.(2001){Rho}, {Jarrett}, {Cutri}, and
  {Reach}}]{Rho(2001)ApJ_547_885}
{Rho}, J., {Jarrett}, T.~H., {Cutri}, R.~M., {Reach}, W.~T., Feb. 2001.
  {Near-Infrared Imaging and [O I] Spectroscopy of IC 443 using Two Micron All
  Sky Survey and Infrared Space Observatory}. ApJ 547, 885--898.

\bibitem[{{Rosenthal} et~al.(2000){Rosenthal}, {Bertoldi}, and
  {Drapatz}}]{Rosenthal(2000)A&A_356_705}
{Rosenthal}, D., {Bertoldi}, F., {Drapatz}, S., Apr. 2000. {ISO-SWS
  observations of OMC-1: H\_2 and fine structure lines}. A\&A 356, 705--723.

\bibitem[{{Shin} et~al.(2008){Shin}, {Stone}, and
  {Snyder}}]{Shin(2008)ApJ_680_336}
{Shin}, M.-S., {Stone}, J.~M., {Snyder}, G.~F., Jun. 2008. {The
  Magnetohydrodynamics of Shock-Cloud Interaction in Three Dimensions}. ApJ
  680, 336--348.

\bibitem[{{Shinn} et~al.(2009){Shinn}, {Koo}, {Burton}, {Lee}, and
  {Moon}}]{Shinn(2009)ApJ_693_1883}
{Shinn}, J.-H., {Koo}, B.-C., {Burton}, M.~G., {Lee}, H.-G., {Moon}, D.-S.,
  Mar. 2009. {Infrared Studies of Molecular Shocks in the Supernova Remnant
  HB21. I. Thermal Admixture of Shocked H$_{2}$ Gas in the North}. ApJ 693,
  1883--1894.

\bibitem[{{Skrutskie et al.}(2006)}]{Skrutskie(2006)AJ_131_1163s}
{Skrutskie et al.}, Feb. 2006. {The Two Micron All Sky Survey (2MASS)}. AJ 131,
  1163--1183.

\bibitem[{{Snow} and {McCall}(2006)}]{Snow(2006)ARA&A_44_367}
{Snow}, T.~P., {McCall}, B.~J., Sep. 2006. {Diffuse Atomic and Molecular
  Clouds}. ARA\&A 44, 367--414.

\bibitem[{{Stetson}(1987)}]{Stetson(1987)PASP_99_191}
{Stetson}, P.~B., Mar. 1987. {DAOPHOT - A computer program for crowded-field
  stellar photometry}. PASP 99, 191--222.

\bibitem[{{Tappe} et~al.(2006){Tappe}, {Rho}, and
  {Reach}}]{Tappe(2006)ApJ_653_267}
{Tappe}, A., {Rho}, J., {Reach}, W.~T., Dec. 2006. {Shock Processing of
  Interstellar Dust and Polycyclic Aromatic Hydrocarbons in the Supernova
  Remnant N132D}. ApJ 653, 267--279.

\bibitem[{{Tatematsu} et~al.(1990){Tatematsu}, {Fukui}, {Landecker}, and
  {Roger}}]{Tatematsu(1990)A&A_237_189}
{Tatematsu}, K., {Fukui}, Y., {Landecker}, T.~L., {Roger}, R.~S., Oct. 1990.
  {The interaction of the supernova remnant HB 21 with the interstellar medium
  - CO, H I, and radio continuum observations}. A\&A 237, 189--200.

\bibitem[{{Thompson}(1973)}]{Thompson(1973)ApJ_181_1039}
{Thompson}, R.~I., May 1973. {Conditions for Carbon Monoxide Vibration-Rotation
  LTE in Late Stars}. ApJ 181, 1039--1054.

\bibitem[{{Tielens}(2008)}]{Tielens(2008)ARA&A_46_289}
{Tielens}, A.~G.~G.~M., Sep. 2008. {Interstellar Polycyclic Aromatic
  Hydrocarbon Molecules}. ARA\&A 46, 289--337.

\bibitem[{{Timmermann}(1998)}]{Timmermann(1998)ApJ_498_246}
{Timmermann}, R., May 1998. {Ortho-H 2/Para-H 2 Ratio in Low-Velocity Shocks}.
  ApJ 498, 246.

\bibitem[{{van Dishoeck}(2004)}]{vanDishoeck(2004)ARA&A_42_119}
{van Dishoeck}, E.~F., Sep. 2004. {ISO Spectroscopy of Gas and Dust: From
  Molecular Clouds to Protoplanetary Disks}. ARA\&A 42, 119--167.

\bibitem[{{Weingartner} and {Draine}(2001)}]{Weingartner(2001)ApJ_548_296}
{Weingartner}, J.~C., {Draine}, B.~T., Feb. 2001. {Dust Grain-Size
  Distributions and Extinction in the Milky Way, Large Magellanic Cloud, and
  Small Magellanic Cloud}. ApJ 548, 296--309.

\bibitem[{{Wilgenbus} et~al.(2000){Wilgenbus}, {Cabrit}, {Pineau des
  For{\^e}ts}, and {Flower}}]{Wilgenbus(2000)A&A_356_1010}
{Wilgenbus}, D., {Cabrit}, S., {Pineau des For{\^e}ts}, G., {Flower}, D.~R.,
  Apr. 2000. {The ortho:para-H\_2 ratio in C- and J-type shocks}. A\&A 356,
  1010--1022.

\bibitem[{{Wilson} et~al.(2003){Wilson}, {Eikenberry}, {Henderson}, {Hayward},
  {Carson}, {Pirger}, {Barry}, {Brandl}, {Houck}, {Fitzgerald}, and
  {Stolberg}}]{Wilson(2003)inproc}
{Wilson}, J.~C., {Eikenberry}, S.~S., {Henderson}, C.~P., {Hayward}, T.~L.,
  {Carson}, J.~C., {Pirger}, B., {Barry}, D.~J., {Brandl}, B.~R., {Houck},
  J.~R., {Fitzgerald}, G.~J., {Stolberg}, T.~M., Mar. 2003. {A Wide-Field
  Infrared Camera for the Palomar 200-inch Telescope}. In: {Iye}, M.,
  {Moorwood}, A.~F.~M. (Eds.), Proceedings of the SPIE. Vol. 4841. Bellingham,
  WA, U.S.A. : SPIE, pp. 451--458.

\end{thebibliography}

\clearpage
\begin{figure}
\caption{
	Regions observed by \akari. 
	The IRC pointing-observation regions (``Cloud N'' and ``Cloud S'') are overlaid as two boxes ($\sim10'\times10'$) on the 1420 MHz radio continuum image of HB 21, obtained by using the synthesis telescope at the Dominian Radio Astrophysical Observatory. 
	The Palomar WIRC observations were performed toward a similar region. 
	In this paper, we present the Cloud S data.
	The Cloud N data were presented in \cite{Shinn(2009)ApJ_693_1883}.
	The 1420 MHz radio continuum image is kindly provided by T. L. Landecker.
} \label{fig-obs}
\end{figure}

\clearpage
\begin{figure}
\caption{
	The \akari{} IRC and Palomar WIRC images of Cloud S. 
	See Table~\ref{tbl-obs} for the band definitions for the IRC images. 
	(\emph{upper-panels}) IRC N3, N4, and S7 band images. 
	(\emph{lower-panels}) IRC S11 band, \CO{} 230.583 GHz \citep{Koo(2001)ApJ_552_175}, and \HH{} 2.122 \um{} images. 
	The peak positions (``S1'' and ``S2''), where broad CO molecular lines were observed, are indicated with a `+' over all images \citep[cf.][]{Koo(2001)ApJ_552_175}. 
	Bright point-sources were masked out, shown by the white circles in the IRC images.
	The center position of the WIRC \HH{} image was mislocated, hence the partial image is shown here.
	A scale bar at the upper-left corner shows $1'$.
} \label{fig-result}
\end{figure}

\clearpage
\begin{figure}
\caption{
	The IRC RGB image (\emph{left}) and the WIRC \HH{} image (\emph{right}) of the cloud S1. 
	The RGB image is composed of S11 (R), S7 (G), and N4 (B) band images, i.e. 11 \um{} + 7 \um{} + 4 \um. 
	All colors scale linearly, and fully cover the dynamic range of the diffuse features.
	The annular regions selected for the intensity measurement are indicated as two concentric white circles.
	The inner circle is the source region and the outer annular region is the background region.
	Circular areas around possible point sources were excluded during the intensity measurement to avoid possible contamination. 
	These areas are indicated as \emph{white circles with a red slash}. 
	Bright point-sources are masked out, and their positions are indicated by white circles with black shading.
	For the comparison between the IRC and WIRC intensity, a tetragonal region is excluded from the IRC images.
	The cross (`+') indicates the peak positions (``S1''), where broad CO molecular lines were observed, as in Figure \ref{fig-result}. 
} \label{fig-rgb}
\end{figure}

\clearpage
\begin{figure}
\center{
\includegraphics[scale=0.5,angle=90]{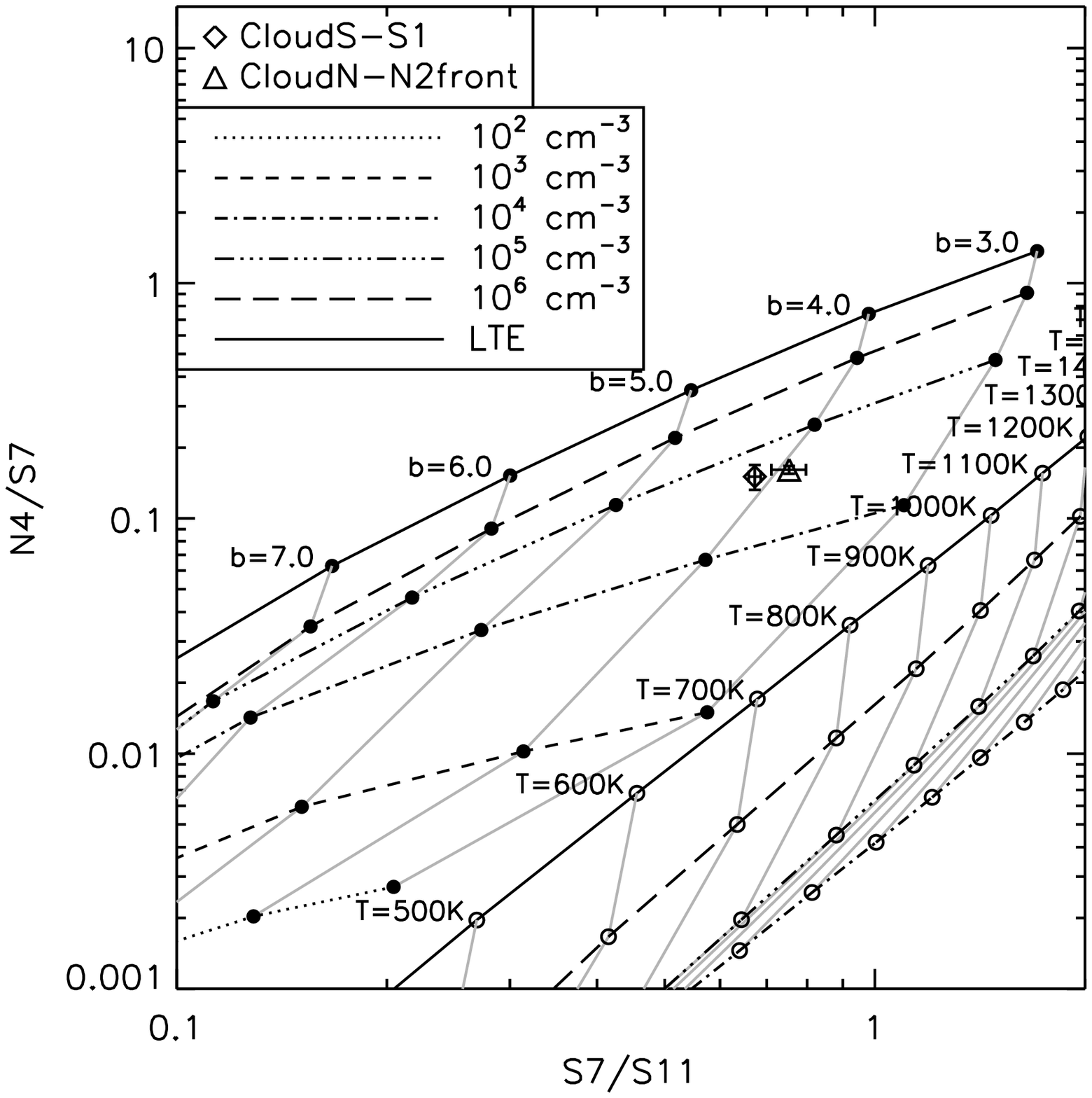}
}
\caption{
	The IRC color-color diagram for the cloud S1, with the colors of the N2front in Cloud N plotted for comparison (cf.~section \ref{dis-ccd}).
	The axes represent the ratio of the intensities in the corresponding IRC bands. 
	The data points are shown by the diamond (S1) and triangle (N2front).
	The expected colors for both isothermal ($\S$\ref{cshock-iso}) and power-law-thermal ($\S$\ref{cshock-pow}) cases are indicated as \emph{open circles} ($\circ$) and \emph{filled circles} ($\bullet$), respectively.
	OPR=3.0 is assumed for both cases.
	The different types of \emph{black} lines connect points of equal \nhh{} and the LTE case. 
	The \emph{grey} solid lines connect points of equal power-law index ($b$) or equal temperature ($T$). 
	The values for the power-law index and temperature are also indicated.
} \label{fig-ccd}
\end{figure}

\clearpage
\begin{figure}
\center{
\includegraphics[scale=0.25,angle=90]{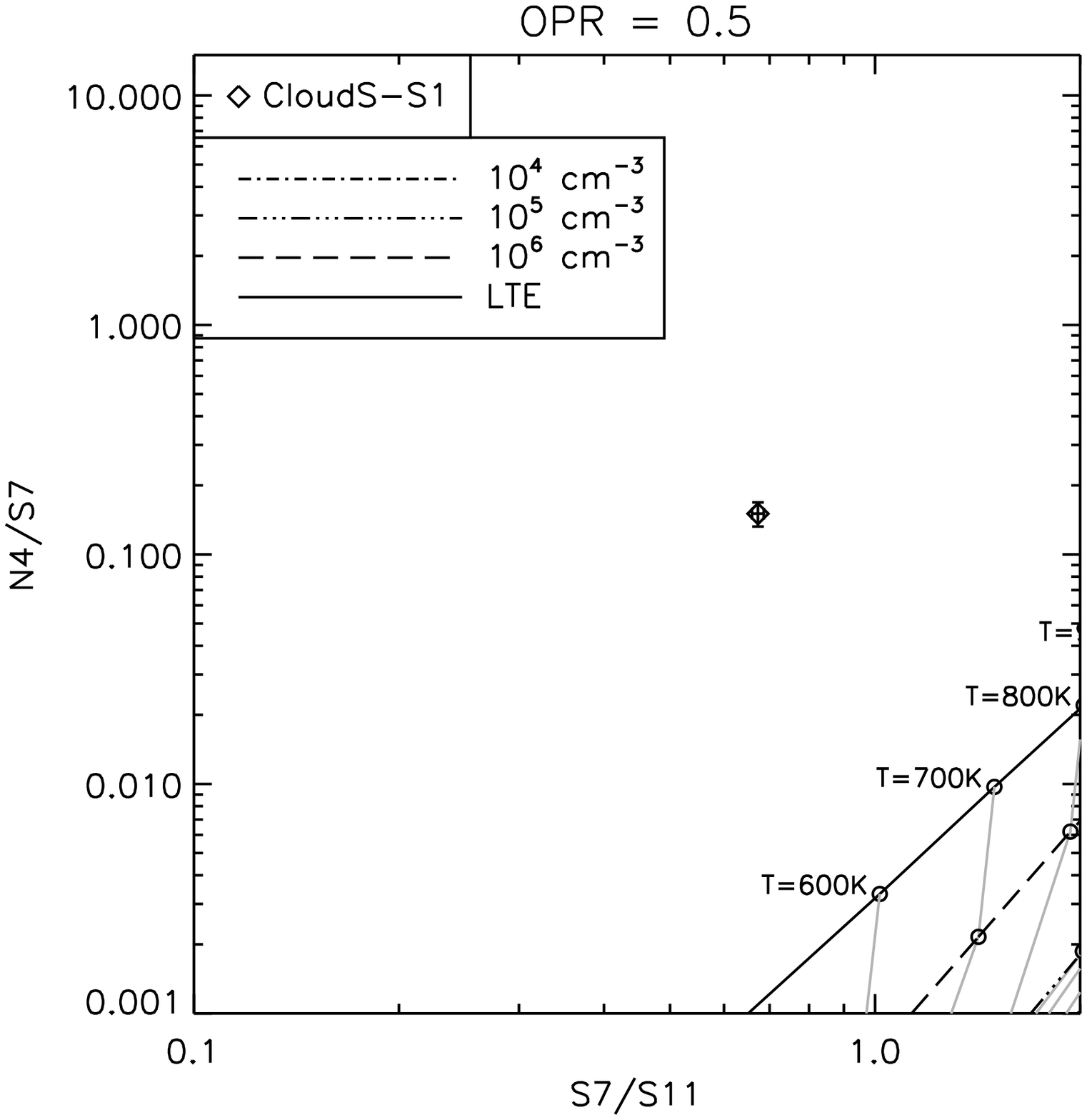}
\includegraphics[scale=0.25,angle=90]{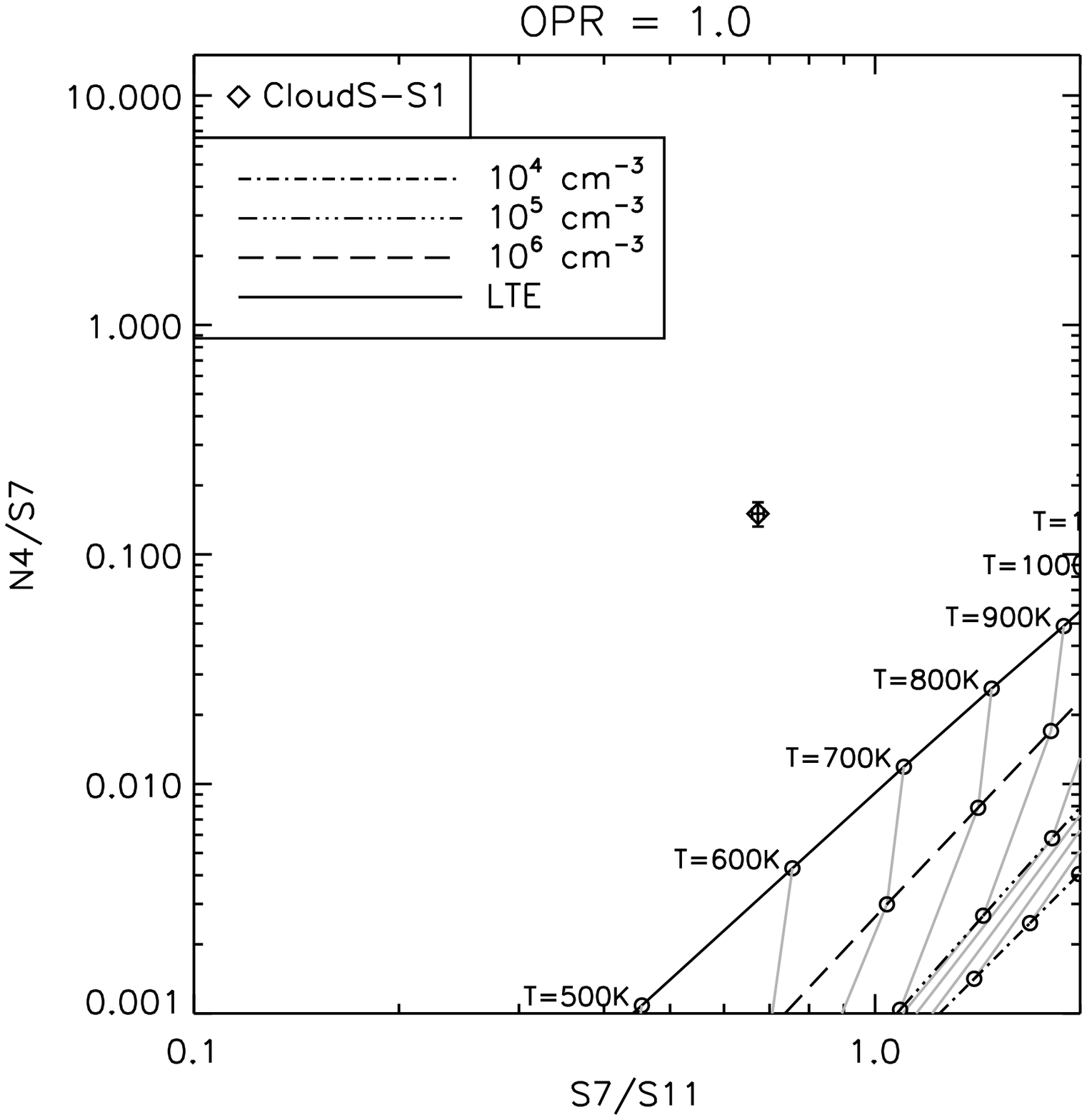}
\includegraphics[scale=0.25,angle=90]{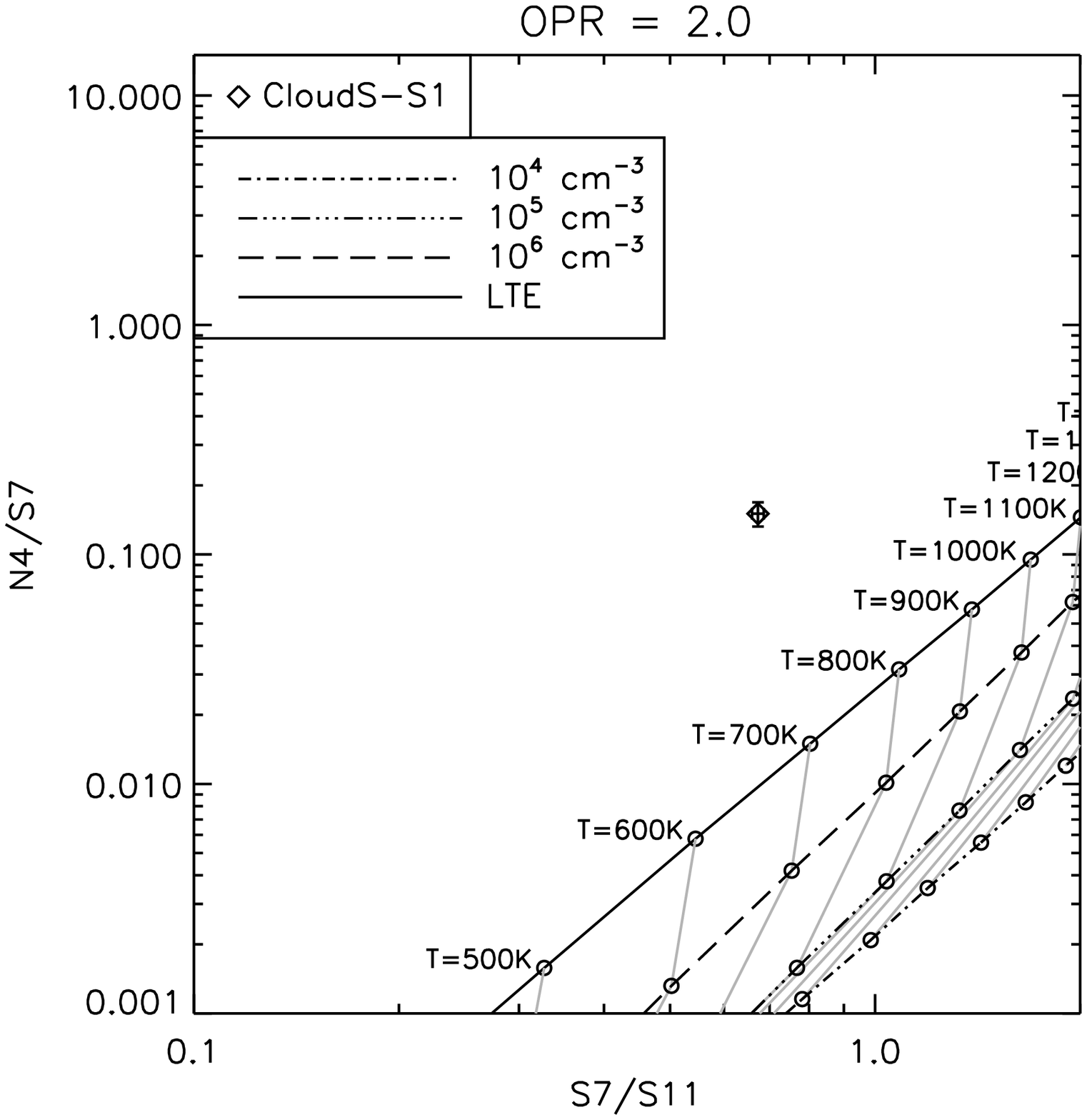}
\includegraphics[scale=0.25,angle=90]{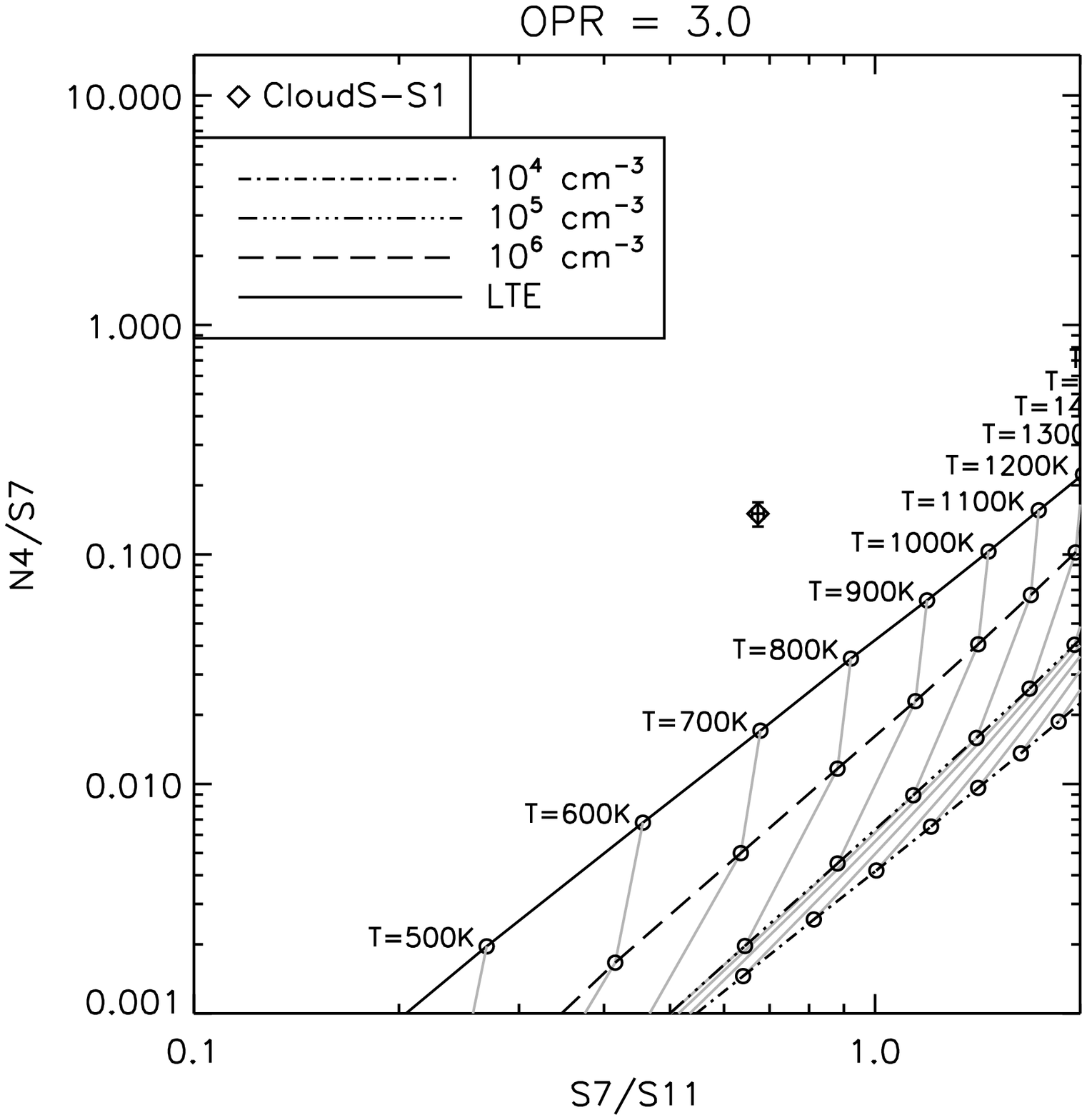}
\includegraphics[scale=0.25,angle=90]{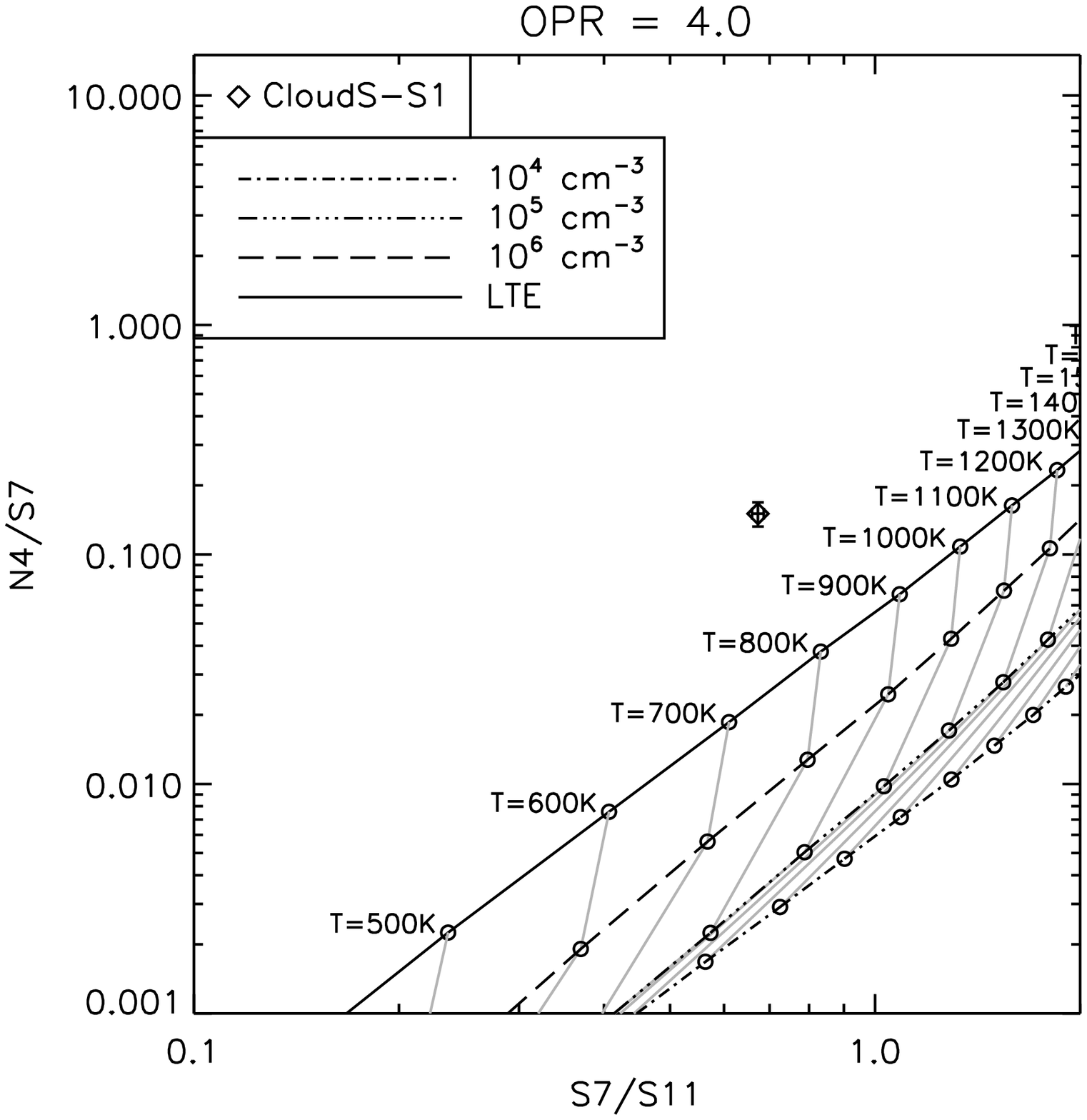}
\includegraphics[scale=0.25,angle=90]{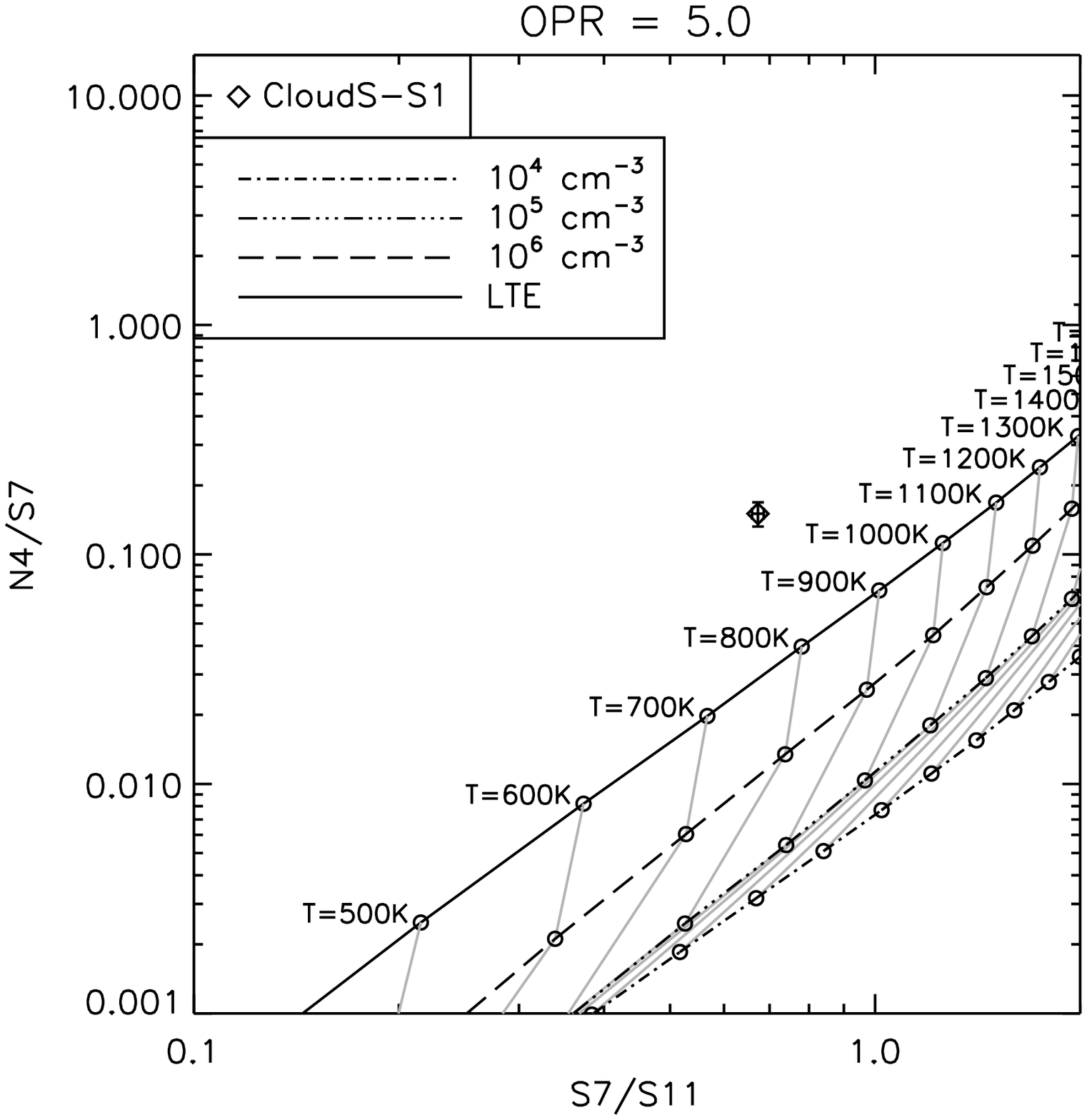}
}
\caption{
	The IRC color-color diagrams for the cloud S1 with the expected colors for \emph{isothermal} cases of various OPRs (cf.~$\S$\ref{cshock-iso}).
	The rest of the graph is the same with Figure~\ref{fig-ccd}.
} \label{fig-opr}
\end{figure}

\clearpage
\begin{figure}
\center{
\includegraphics[scale=0.5,angle=90]{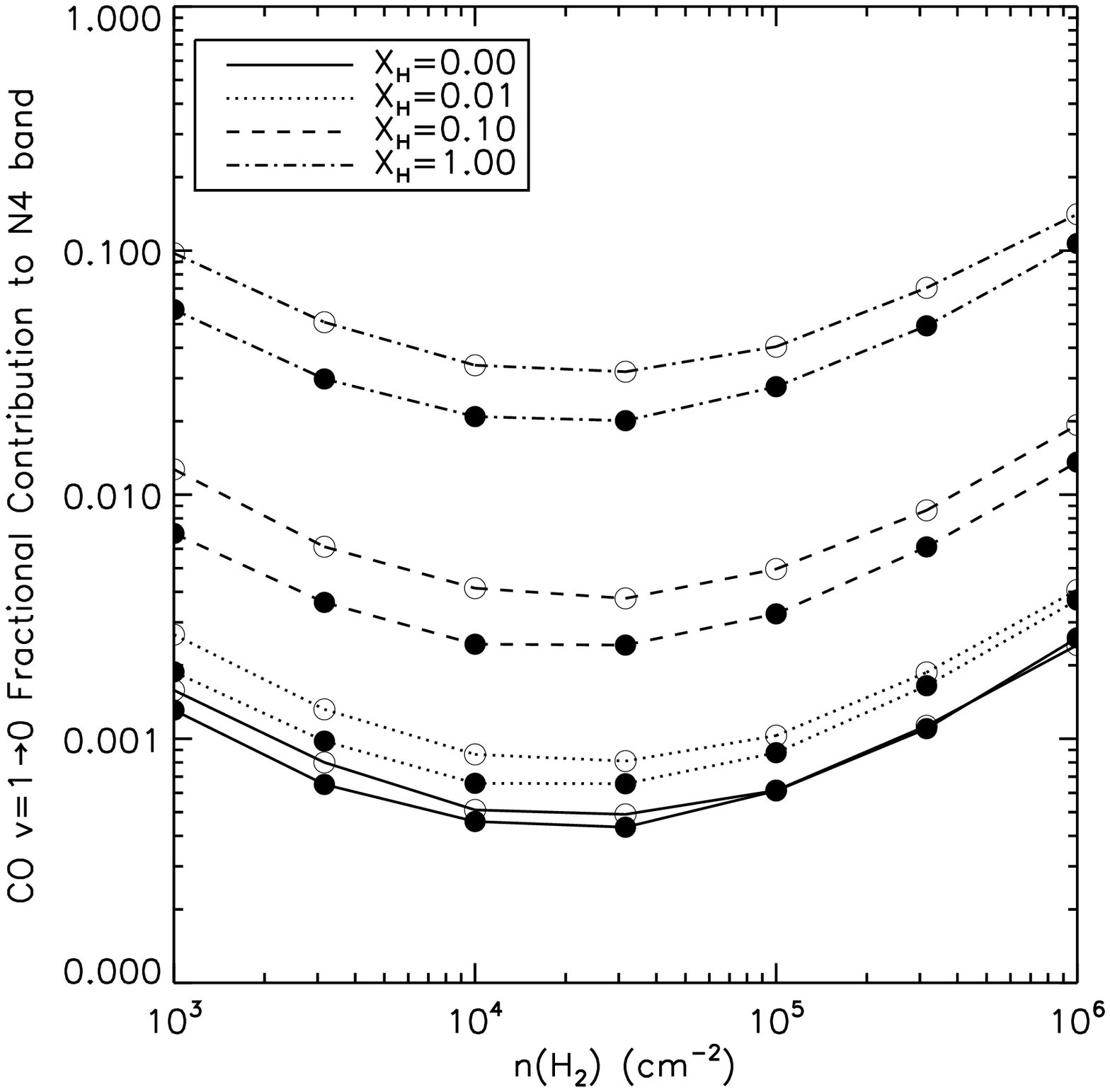}
}
\caption{
	Fractional contribution of \COv{} emission lines to the IRC N4 band as a function of \nhh.
	\emph{Filled circles} and \emph{open circles} are for $b=4.0$ and $b=5.0$, respectively.
	$X_H$ is the fractional abundance of atomic hydrogen to molecular hydrogen, $N$(\Hi)/\Nhh.
	The fractional abundance of CO to \hh{} is assumed to be $10^{-4}$.
	Overall, in the range of \nhh=$10^3-10^6$ \ncm, the fractional contribution of \COv{} to the IRC N4 band is less than 0.1.
} \label{fig-cov}
\end{figure}

\clearpage
\begin{figure}
\center{
\includegraphics[scale=0.33,angle=90]{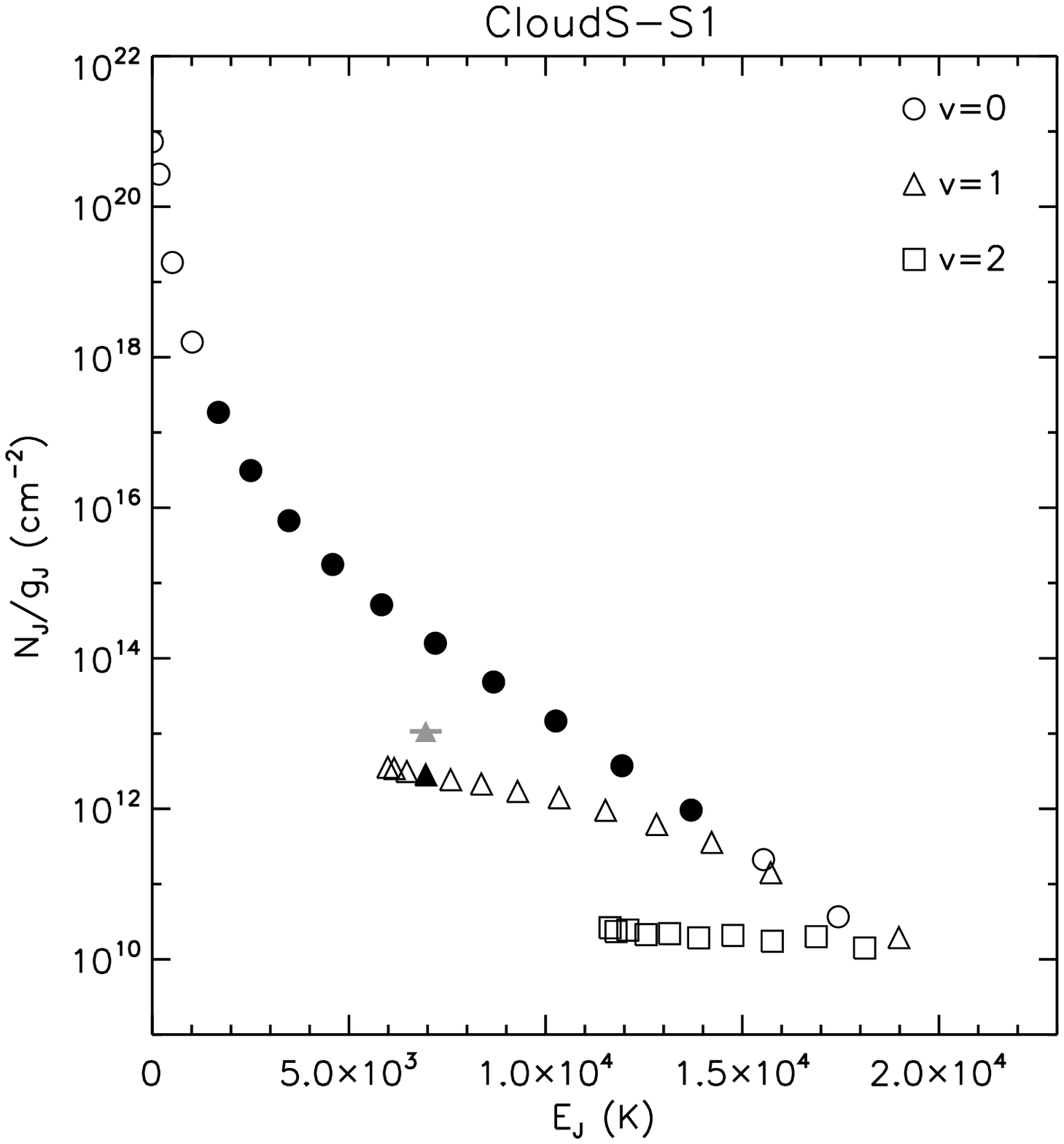}
\includegraphics[scale=0.33,angle=90]{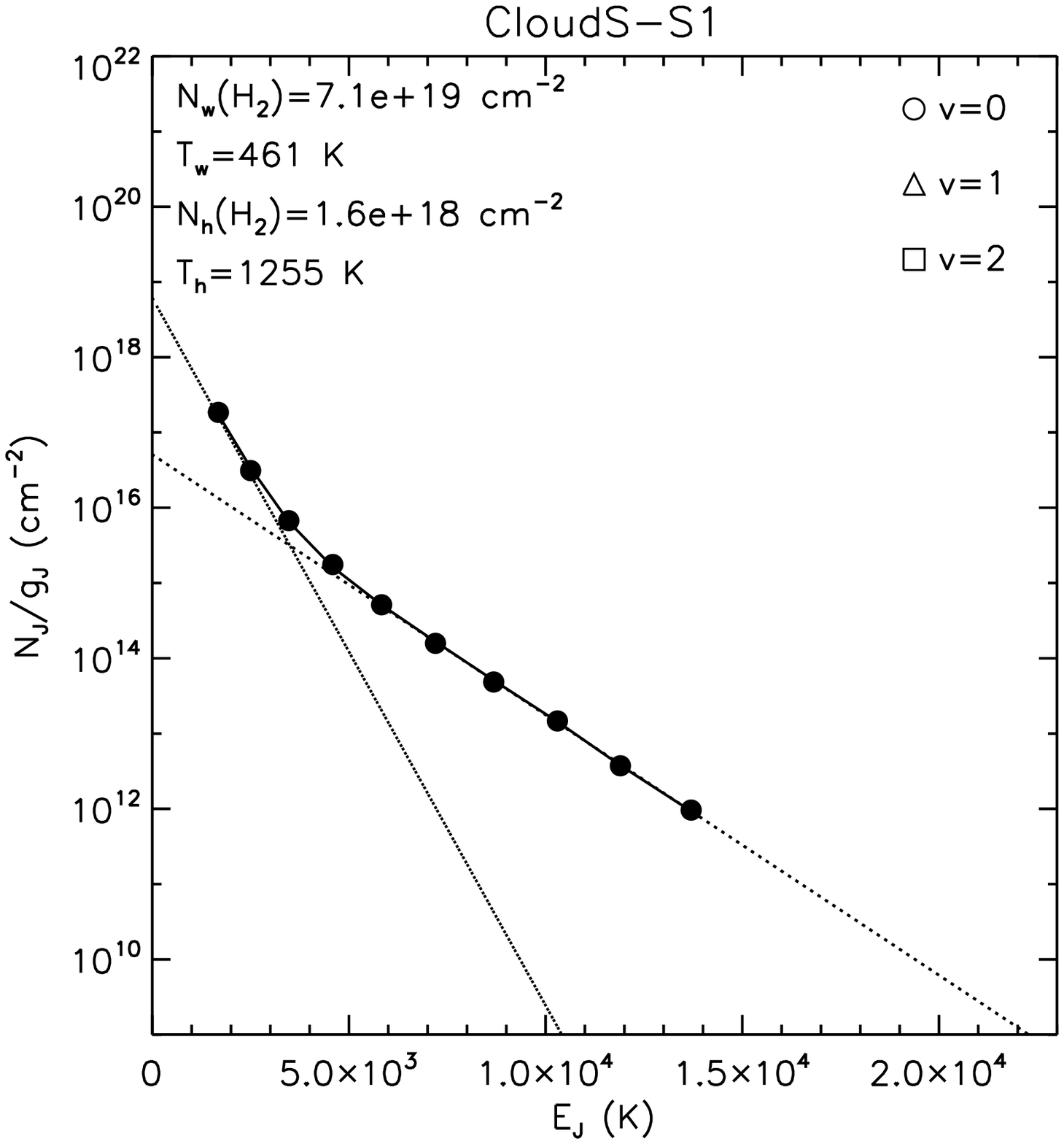}
\includegraphics[scale=0.33,angle=90]{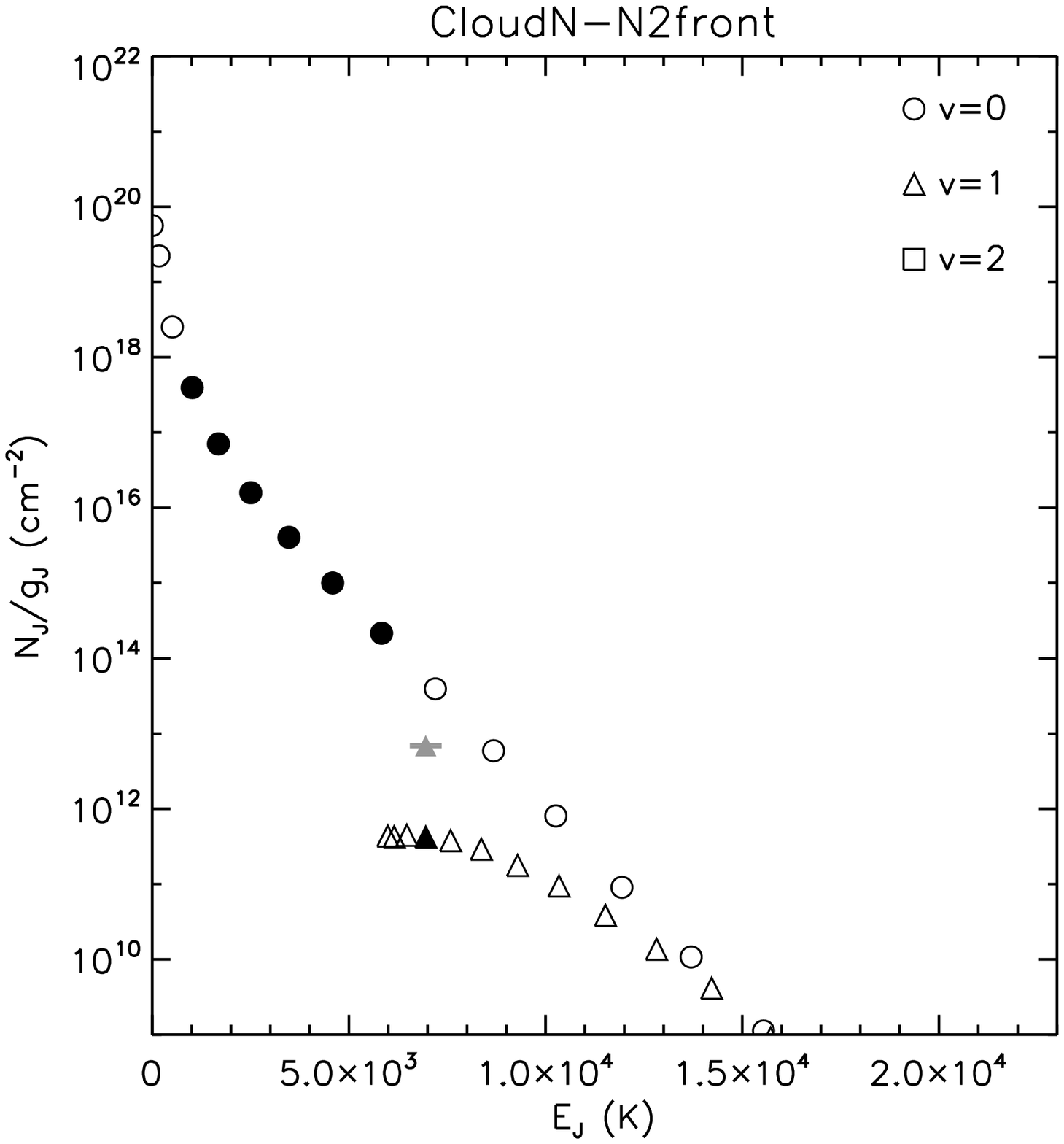}
\includegraphics[scale=0.33,angle=90]{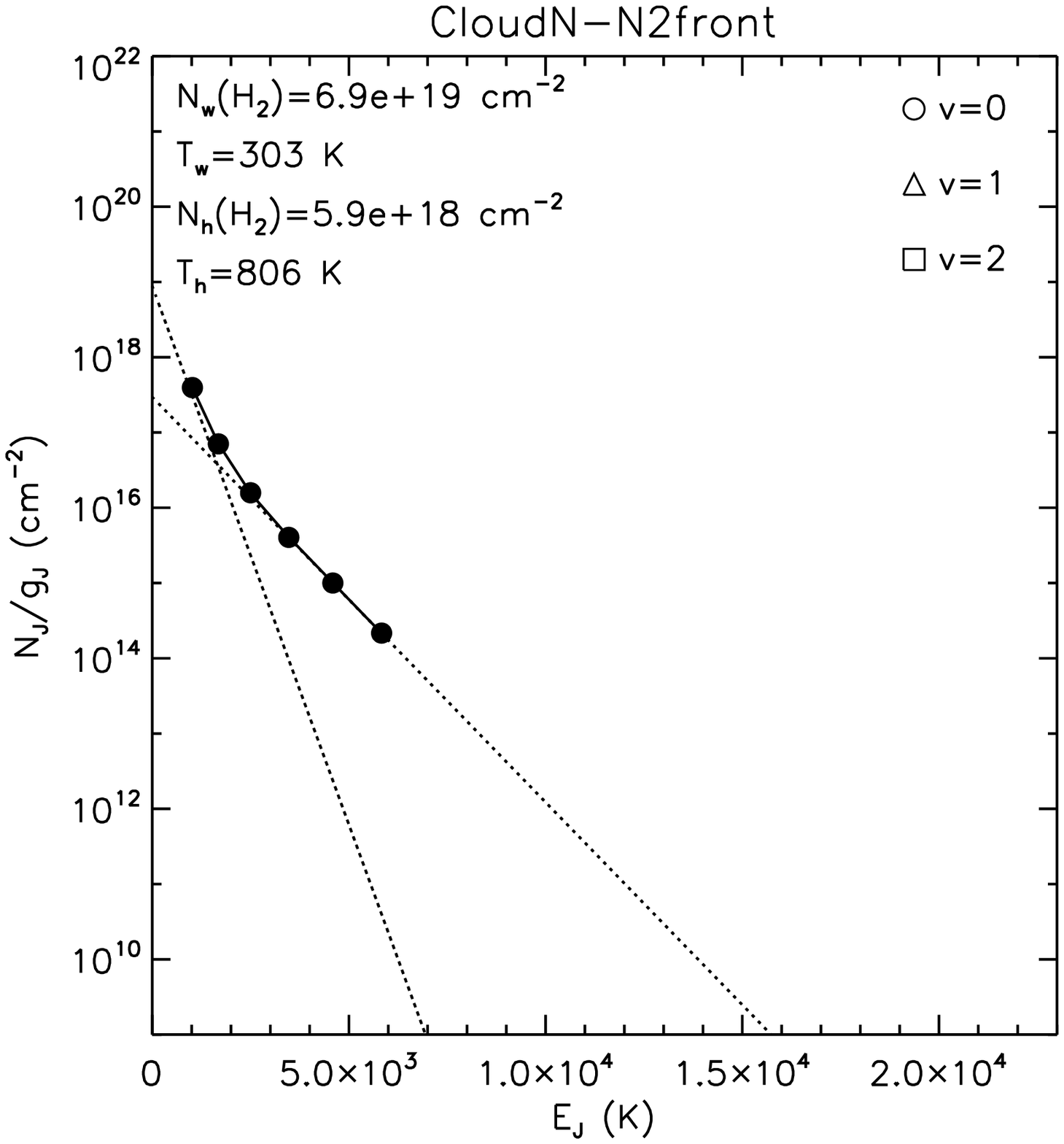}
}
\caption{
	(\emph{top-left}) The population diagram of the cloud S1, derived from the IRC color-color diagram (Fig.~\ref{fig-ccd}) using the power-law admixture model (cf. $\S$ \ref{cshock-pow}).
	The vibrational levels of $\upsilon=0,1,2$ are designated by circles, triangles, and squares, respectively.
	The circles show the ``ankle-like'' curvature, which turn slightly upward around $E_{J}\sim3000$ K as $E_J$ increases.
	The filled circles are the levels which contribute to the corresponding IRC bands (cf.~Table \ref{tbl-cont}).
	The filled triangle is the level of ($\upsilon, J$)=(1,3), the upper level of the emission line \HH.
	The grey filled triangle with an error bar is the population of ($\upsilon, J$)=(1,3) derived from the observed \HH{} intensity, extinction corrected.
	(\emph{top-right}) Two temperature LTE fitting of the \hh{} levels of the cloud S1, which contribute to the corresponding IRC bands (the filled circles of the \emph{left} figure).
	OPR=3.0 is adopted.
	The obtained fitting parameters are ($T_w,N_w$)=(461 K, $7.1\times10^{19}$ \Ncm) and ($T_h,N_h$)=(1225 K, $1.6\times10^{18}$ \Ncm).
	See the section \ref{dis-col} for a further description.
	(\emph{bottom-panels}) The same plots for the N2front in Cloud N (Paper I).
} \label{fig-pop}
\end{figure}

\clearpage
\begin{figure}
\center{
\includegraphics[scale=0.5,angle=90]{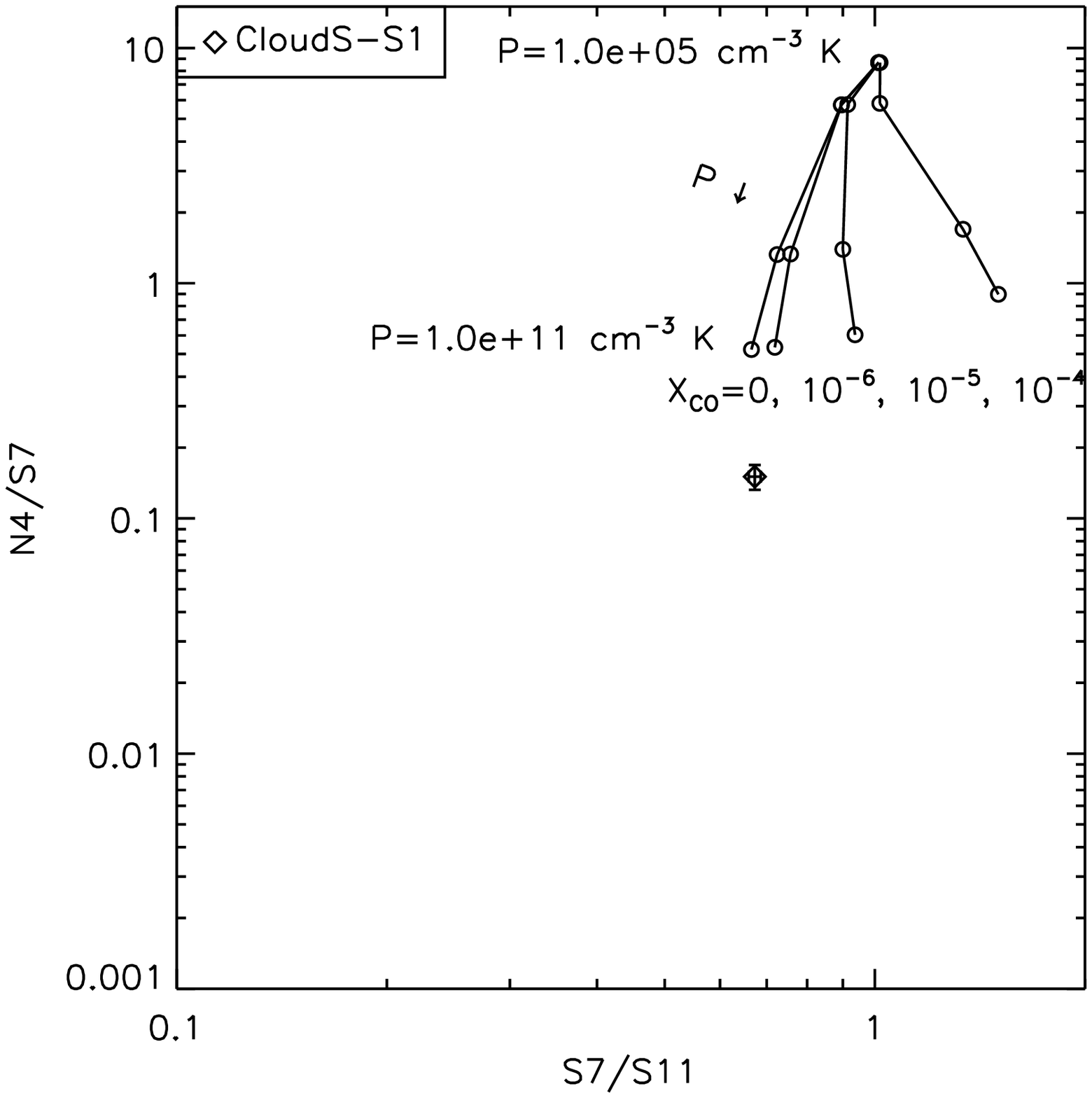}
}
\caption{
	The expected IRC colors for a partially dissociative J-shock model \citep{Brand(1988)ApJ_334_L103,Burton(1989)inproca}. 
	The axes represent the ratio of the intensities in the corresponding IRC bands. 
	The data point is shown by the diamond.
	The connected \emph{open-circles} have the same fractional CO abundance to \Nhh, \Xco; the four lines correspond to \Xco$=0, 10^{-6}, 10^{-5}, 10^{-4}$, from left to right. 
	The pressure increases from $10^5$ \ncmK{} to $10^{11}$ \ncmK{}, by a factor of $10^2$, along each line. 
} \label{fig-other}
\end{figure}

\clearpage
\begin{figure}
\caption{
	(\emph{top}) The schematic description for the bow shock picture
	(\emph{bottom}) The schematic description for the shocked clumpy ISM picture
} \label{fig-pic}
\end{figure}

\clearpage
\begin{figure}
\center{
\includegraphics[scale=0.25,angle=90]{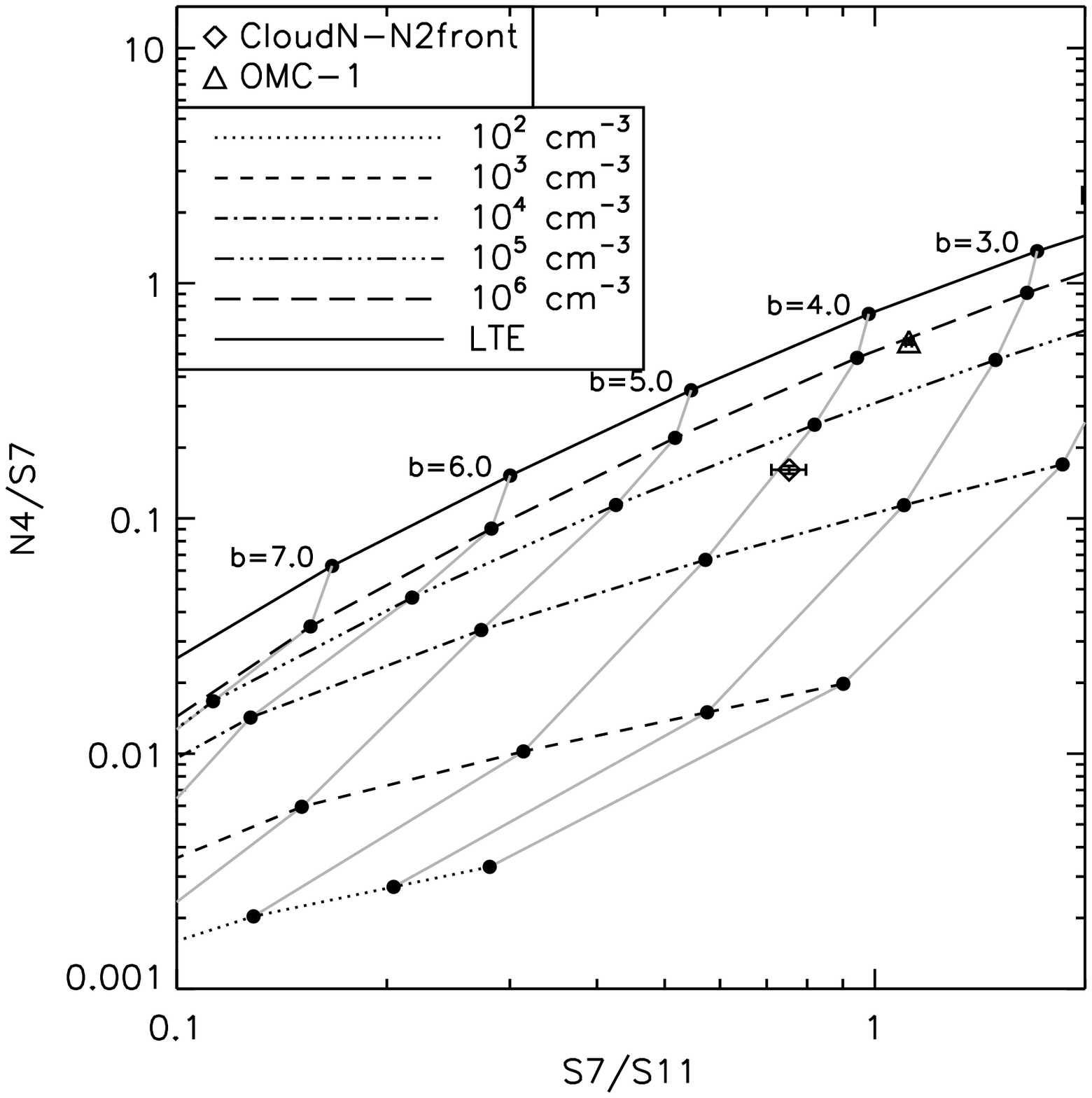}
\includegraphics[scale=0.25,angle=90]{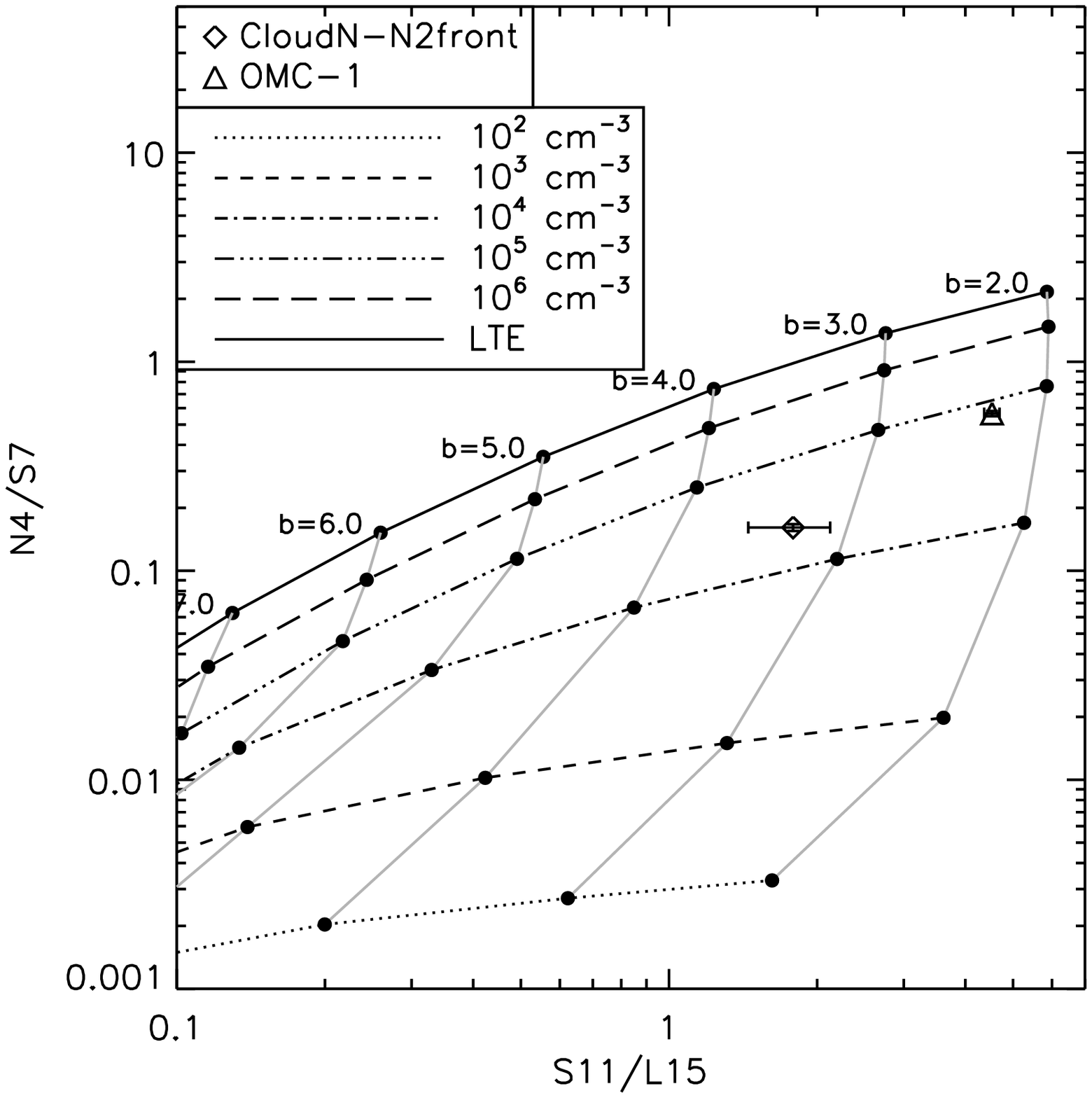}
\includegraphics[scale=0.25,angle=90]{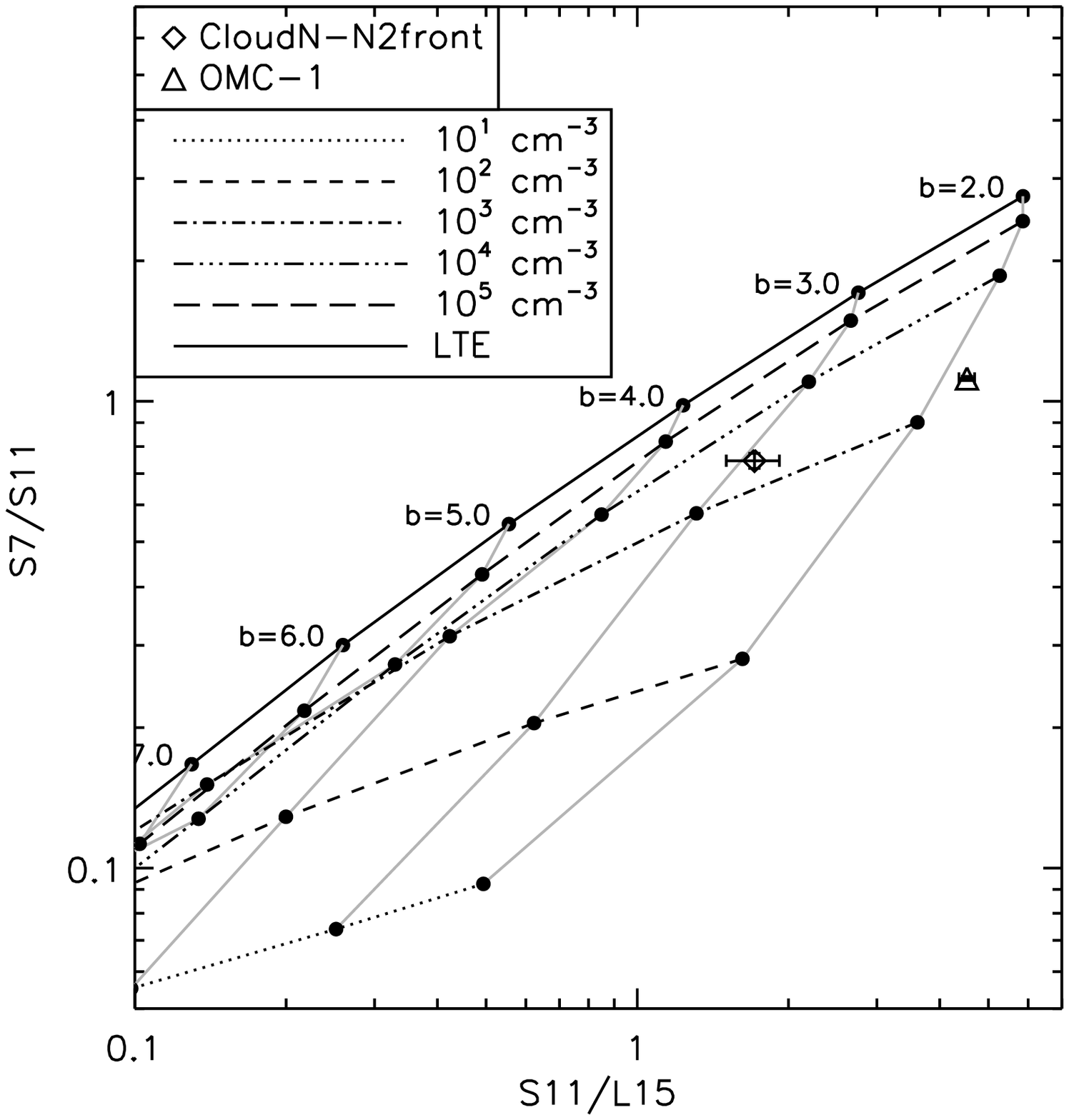}
}
\caption{
	The IRC color-color diagram for N2front and Orion Molecular Cloud-1: (\emph{leftmost}) N4/S7 versus S7/S11, (\emph{middle}) N4/S7 versus S11/L15, and (\emph{rightmost}) S7/S11 versus S11/L15.
	The grids are the expected colors from the thermal admixture model (cf. section \ref{cshock-pow}).
	Depending on which IRC bands are used for the color-color diagram, the obtained parameters, $b$ and \nhh, are different for each data point; the shorter-wavelength bands return the higher-$b$ and higher-\nhh{} (cf.~N4/S7 vs. S7/S11 and S7/S11 vs. S11/L15 diagrams).
	See section \ref{dis-ccd} for detail.
}\label{fig-n2f}
\end{figure}

\clearpage
\begin{table}
\caption{Summary of the \akari{} IRC Observations \label{tbl-obs}}
\begin{tabular}{ccccc}
\hline
Channel & Filter & Wavelength & Imaging & Data ID \\
& & coverage$^{a}$ & Resolution ($\Gamma$) \\
(pixel size) &  & (\um) & (FWHM, $''$) \\
\hline
NIR         &   N3 &    2.7--3.8 & 4.0 & 1402804\\
($1.46''\times1.46''$)         &   N4 &    3.6--5.3 & 4.2 & 1402804\\
\hline
MIR-S       &   S7 &     5.9--8.4 & 5.1 & 1402804\\
($2.34''\times2.34''$)       &   S11 &     8.5--13.1 & 4.8 & 1402804\\
\hline
\end{tabular}\\
{$^a$ Defined as where the responsivity is larger than $1/e$ of the peak \\for the imaging mode. See \cite{Onaka(2007)PASJ_59_S401s}.}
\end{table}

\begin{table}
{\scriptsize
\caption{Results toward Cloud S \label{tbl-result}}
\begin{tabular}{ccccccc}
\hline
{Region} &  {N4} & {S7} & {S11} & {N4/S7} & {S7/S11} & {\HH{}{$^a$}} \\
&(MJy sr$^{-1}$) &(MJy sr$^{-1}$) &(MJy sr$^{-1}$) & & &(\luerg)\\
\hline
S1                                 &      0.16$\pm$0.02&      1.05$\pm$0.01&      1.56$\pm$0.03&      0.15$\pm$0.02&      0.67$\pm$0.01&         (5.9$\pm$0.2)$\times10^{-6}$\\

\hline
\end{tabular}\\
{$^a$ Extinction-corrected intensity with $N$(H)=$3.5\times10^{21}$ \Ncm{} ($A_V=1.8$ mag for $R_V$=3.1). See text for detail.}
}
\end{table}

\begin{table}
{
\caption{Derived Parameters for the Power-law Admixture Model{$^a$} and the Predicted \HH{} Intensity \label{tbl-par}}
\begin{tabular}{ccccc}
\hline
{Region} & {$n$(H$_2$)} & {$b$} & {\NhhIRC} & predicted \HH{}\\
&(cm$^{-3}$) & &(\Ncm) &(\luerg) \\
\hline
S1                          &(3.9$_{-1.2}^{+2.1}$)$\times10^{4}$&4.2$_{-0.1}^{+0.1}$&(2.8$_{-0.5}^{+0.2}$)$\times10^{21}$&(1.5$_{-0.3}^{+0.5}$)$\times10^{-6}$\\

\hline
\end{tabular}\\
{$^a$See section \ref{cshock-pow} for the detailed model description.}
}
\end{table}

\begin{table}
{
\caption{Derived Contribution of \hh{} line emission to the IRC bands \label{tbl-cont}}
\begin{tabular}{ccccccc}
\hline
{Transition} & {Wavelength} & {Upper State Energy} & {IRC} & {Weight{$^a$}} & {\% Contribution{$^b$}} \\
&($\mu m$) &(K)\\
\hline
H$_{2}$ $v=0-0$ $S(11)$ &     4.181 &13703 &N4 &     0.362 &7\\
H$_{2}$ $v=0-0$ $S(10)$ &     4.410 &11940 &N4 &     0.372 &6\\
\COv{} &     4.662 &3086 &N4 &     0.390 &see $\S$\ref{cshock-pow}\\
H$_{2}$ $v=0-0$ $S(9)$ &     4.695 &10261 &N4 &     0.388 &42\\
H$_{2}$ $v=0-0$ $S(8)$ &     5.053 &8677 &N4 &     0.285 &19\\
H$_{2}$ $v=0-0$ $S(7)$ &     5.511 &7197 &N4 &     0.070 &24\\
H$_{2}$ $v=0-0$ $S(6)$ &     6.109 &5830 &S7 &     0.346 &8\\
H$_{2}$ $v=0-0$ $S(5)$ &     6.909 &4586 &S7 &     0.530 &52\\
H$_{2}$ $v=0-0$ $S(4)$ &     8.026 &3474 &S7 &     0.961 &39\\
H$_{2}$ $v=0-0$ $S(3)$ &     9.665 &2504 &S11 &     0.921 &79\\
H$_{2}$ $v=0-0$ $S(2)$ &    12.279 &1682 &S11 &     0.610 &20\\
\hline
\end{tabular}\\
	{
	$^a$ 
	In units of 10$^4$ MJy sr$^{-1}$/(\luerg).
	See the text for the description.\\
	$^b$
	The contributions are given to the nearest integer. 
	Hence, their sum can be less than 100\%.
	}
}
\end{table}


\begin{table}
{
\caption{New Intensity Determination for N2front in Cloud N \label{tbl-n2f}}
\begin{tabular}{ccccc}
\hline
{Region} & {N4} & {S7} & {S11} & {L15} \\
& (MJy sr$^{-1}$) & (MJy sr$^{-1}$) & (MJy sr$^{-1}$) & (MJy sr$^{-1}$)\\
\hline
N2front                     &     0.108$\pm$0.003&      0.67$\pm$0.02&      0.89$\pm$0.04&      0.50$\pm$0.09\\
\hline
\end{tabular}\\
}
\end{table}

\end{document}